\documentclass[english,numbers, numbers, sort&compress]{IEEEtran}
\usepackage[T1]{fontenc}
\usepackage[latin9]{inputenc}
\usepackage{float}
\usepackage{textcomp}
\usepackage{amsthm}
\usepackage{amsmath}
\usepackage{amssymb}
\usepackage{graphicx}
\usepackage{esint}
\usepackage[numbers]{natbib}

\makeatletter

\providecommand{\tabularnewline}{\\}
\floatstyle{ruled}
\newfloat{algorithm}{tbp}{loa}
\providecommand{\algorithmname}{Algorithm}
\floatname{algorithm}{\protect\algorithmname}

  \theoremstyle{remark}
  \newtheorem{rem}{\protect\remarkname}
  \theoremstyle{plain}
  \newtheorem{lem}{\protect\lemmaname}
 \ifx\proof\undefined\
   \newenvironment{proof}[1][\proofname]{\par
     \normalfont\topsep6\p@\@plus6\p@\relax
     \trivlist
     \itemindent\parindent
     \item[\hskip\labelsep
           \scshape
       #1]\ignorespaces
   }{%
     \endtrivlist\@endpefalse
   }
   \providecommand{\proofname}{Proof}
 \fi
  \theoremstyle{plain}
  \newtheorem{thm}{\protect\theoremname}
  \theoremstyle{plain}
  \newtheorem{cor}{\protect\corollaryname}

\IEEEoverridecommandlockouts
\usepackage{graphicx}

\def\@tempa{12}\ifx\@ptsize\@tempa
\def\@normalsize{\@setsize\normalsize{14pt}\xiipt\@xiipt
\abovedisplayskip 2pt plus2pt minus2pt\belowdisplayskip \abovedisplayskip
\abovedisplayshortskip \z@ plus3pt\belowdisplayshortskip .2pt plus2pt minus2pt}
\def\small{\@setsize\small{11.4pt}\xpt\@xpt}
\def\footnotesize{\@setsize\footnotesize{10pt}\ixpt\@ixpt}
\def\scriptsize{\@setsize\scriptsize{9pt}\viiipt\@viiipt}
\def\tiny{\@setsize\tiny{8pt}\vipt\@vipt}
\def\large{\@setsize\large{18pt}\xivpt\@xivpt}
\def\Large{\@setsize\Large{22pt}\xviipt\@xviipt}
\def\LARGE{\@setsize\LARGE{25pt}\xxpt\@xxpt}
\def\huge{\@setsize\huge{30pt}\xxvpt\@xxvpt}

\fi

\usepackage{xspace}
\usepackage{fixfoot}
\DeclareFixedFootnote*{\FeasibleFootnote}{Particularly, if those equations do not have feasible solutions, it means that some relays are with too poor relay-destination channels, and even if the source uses an infinite transmit power, the required utility still cannot be satisfied.}

\@ifundefined{showcaptionsetup}{}{%
 \PassOptionsToPackage{caption=false}{subfig}}
\usepackage{subfig}
\makeatother

\usepackage{babel}
\providecommand{\corollaryname}{Corollary}
\providecommand{\lemmaname}{Lemma}
\providecommand{\remarkname}{Remark}
\providecommand{\theoremname}{Theorem}

\begin{document}

\title{Transmit Power Minimization for Wireless Networks with Energy Harvesting
Relays}

\author{Yaming Luo, Jun Zhang, \emph{Senior Member,} \emph{IEEE,} and Khaled
B. Letaief, \textit{Fellow, IEEE}%
\thanks{This work is supported by the Hong Kong Research Grant Council under
Grant No. 610212. 

The authors are with the Dept. of Electronic and Computer Engineering,
Hong Kong University of Science and Technology, Hong Kong. Email:
\{luoymhk, eejzhang, eekhaled\}@ust.hk. Khaled B. Letaief is also
with Hamad Bin Khalifa University, Qatar (kletaief@hbku.edu.qa).%
}}
\maketitle
\begin{abstract}
Energy harvesting (EH) has recently emerged as a key technology for
green communications as it can power wireless networks with renewable
energy sources. However, directly replacing the conventional non-EH
transmitters by EH nodes will be a challenge. In this paper, we propose
to deploy extra EH nodes as relays over an existing non-EH network.
Specifically, the considered non-EH network consists of multiple source-destination
(S-D) pairs. The deployed EH relays will take turns to assist each
S-D pair, and \emph{energy diversity} can be achieved to combat the
low EH rate of each EH relay. To make the best of these EH relays,
with the source transmit power minimization as the design objective,
we formulate a joint power assignment and relay selection problem,
which, however, is NP-hard. We thus propose a general framework to
develop efficient sub-optimal algorithms, which is mainly based on
a sufficient condition for the feasibility of the optimization problem.
This condition yields useful design insights and also reveals an \emph{energy
hardening effect}, which provides the possibility to exempt the requirement
of non-causal EH information. Simulation results will show that the
proposed cooperation strategy can achieve near-optimal performance
and provide significant power savings. Compared to the greedy cooperation
method that only optimizes the performance of the current transmission
block, the proposed strategy can achieve the same performance with
much fewer relays, and the performance gap increases with the number
of S-D pairs.\end{abstract}


\section{Introduction}

Wireless communications is evolving towards a green paradigm with
significantly improved energy efficiency and environmental friendliness
\citep{Green_cellular}. In particular, energy harvesting (EH) technology
has recently emerged as a promising approach to power wireless networks
with renewable energy sources \citep{EH_communication_survey,EH_SmallCell_Mag}.
The EH node can obtain energy from the environment \citep{EH_WSN},
including solar energy, thermoelectric energy, vibration energy, RF
energy, etc. As the harvested energy is clean and renewable, environmental
friendliness is innate in EH networks. Besides this, by exempting
the manual battery replacement, EH technology also qualifies the wireless
networks with the ability of virtually perpetual operation \citep{EH_longlive,EH_perpetual}.
These advantages render EH technology a good candidate for green communications. 

A clean-slate design of a wireless network solely powered by energy
harvesting will, however, be challenging, if not impossible. Typically
the harvested energy is in a small amount \citep{EH_survey} and is
also spatially-temporally varying \citep{EH_experiment_solar}. Hence,
it will be difficult to provide satisfactory communication performance
in an EH network. Meanwhile, communications protocols need to be re-designed
to handle the random and intermittent energy arrivals. On the other
hand, considering the implementation cost and efficiency, the existing
communication infrastructure should be fully utilized. Therefore,
it is more realistic to incrementally incorporate the energy harvesting
feature into existing non-EH wireless networks. In this paper, we
will propose to deploy multiple EH relays to assist the transmission
of multiple communication pairs in an existing network, so that the
conventional power consumption of the source nodes can be significantly
reduced. For fairness consideration, in this paper, we consider the
maximum power consumption minimization problem, similar to the minimum
rate maximization problem for throughput-oriented optimization with
fairness consideration \citep{PA_multi_user_relay}. If the transmitters
are mobile devices powered by batteries, this kind of optimization
is directly related to the extension of the network lifetime, which
is of particular importance for many applications such as wireless
sensor networks \citep{Lifetime_WN}. The optimization of network
operations in such a network will require new cooperation strategies,
which form the main focus of this paper.

\subsection{Related Works }

The potential of EH technology has recently spurred a lot of research
activities in the wireless communications community. There are two
main design approaches, i.e. offline and online policies, depending
on non-causal and causal side information (of the energy state or
the channel state), respectively. The offline transmission policies
were investigated in \citep{EH_DWF_time,EH_Broad_time,EH_DWF_Mac,EH_interference,EH_train}
for different kinds of networks, ranging from the point-to-point channel,
broadcast channel, multiple access channel, to interference channel.
The online algorithms, mainly for the point-to-point channel, were
designed in \citep{Energy_allo_EH_cons,EH_DWF_channel,Outage_min_EH_DP,EH_Lyap_HES}.
These studies have revealed the main challenges in designing EH networks,
which are caused by the energy causality constraint, i.e. the energy
consumed so far cannot exceed the total harvested energy. Subsequently,
communication protocols need to be revisited in EH networks. For example,
even when the channel remains unchanged, the transmit power should
still adapt to the random energy arrivals \citep{EH_DWF_channel}.
Moreover, the communication performance is fundamentally limited by
the small-amount and time-varying energy arrivals.

Cooperative communications has been proven to be an effective technique
to improve the communication performance as well as the energy efficiency
for wireless networks \citep{Cooperative_user_diversity,Cooperative_Diversity,Coop_Unif_Cross,Coop_Cog,EH_MultiUser_AFMIMO}.
In particular, its potential in EH wireless networks has been recently
investigated. For single-relay two-hop EH networks, the authors in
\citep{EH_two_hop1,EH_relay_gaussian_Rui,EH_two_hop2,EH_two_hop5,EH_two_hop6,EH_two_hop4,EH_link_Relay}
investigated the problem of power allocation and source/relay transmission
scheduling, and verified that the existing transmission policies for
conventional non-EH communication systems perform poorly. With multiple
EH relays, it turns out that the cooperation strategy design is very
challenging due to the EH constraints. In \citep{EH_RS}, the SER
performance of a multi-relay EH network using a simple relay selection
scheme based on the instantaneous side information of each relay was
analyzed thoroughly. In particular, the coupling effect among different
relays was observed and discussed. Joint power allocation and relay
selection was considered in \citep{EH_RS_PA_ranj} with either non-causal
or causal channel and energy side information, but the complexity
of the proposed algorithms is high. By adopting a fixed transmit power,
an efficient but sub-optimal relay selection method based on the so-called
relative throughput gain was proposed in \citep{EH_RS_LYM}.

All the relay selection methods in \citep{EH_RS,EH_RS_PA_ranj,EH_RS_LYM}
require the channel side information of each relay in each transmission
block, which is challenging to obtain, especially given the low EH
rates of the EH relays. Previous works on EH networks focused on the
cumulative performance over a relatively long duration, while the
short-term performance during a particular transmission block may
not satisfy the application requirement. This short-term performance
is normally a special design obstacle we need to overcome for EH systems,
due to the time-varying EH rate at each EH node. Moreover, only a
single source-destination (S-D) pair has been considered in the previous
studies of EH relaying networks. With multiple S-D pairs, besides
the dynamic energy arrivals at each EH relay, the spatial distribution
of all the EH relays and S-D pairs will sharpen the design challenges
for cooperation strategies.

In this paper, we aim to achieve power conservation for non-EH wireless
networks composed of multiple S-D pairs by deploying multiple EH relays.
For this system, the cooperation strategies in previous works are
no longer applicable. To be practical, we assume that each source
node does not have instantaneous channel state information. To guarantee
satisfactory performance of all the S-D pairs, for each transmission
block each source needs to transmit with the assistance of relays.
Meanwhile, a certain QoS requirement is used to regulate the communication
performance. Note that such a network with multiple S-D pairs is very
generic in practice \citep{PA_multi_user_relay}, especially for ad-hoc
networks and mesh networks.

\subsection{Contributions }

In this paper, we propose an effective and low-complexity cooperation
strategy to reduce the transmit power of multiple S-D pairs with the
help of EH relays. To combat the low EH rate of the relays, \emph{energy
diversity} will be achieved by allowing relays to take turns to forward
the source information. The main contributions are summarized as follows.
\begin{itemize}
\item To optimize the proposed cooperation strategy, we will formulate a
joint power assignment and relay selection problem. The design objective
is to minimize the maximum transmit power among all the S-D pairs.
This problem is found to be NP-hard, and we will propose a general
framework to develop efficient sub-optimal algorithms. The key step
is to derive an easy-to-check sufficient condition for the feasibility
of the optimization problem. The proposed algorithms are of low complexity,
and the optimality can be achieved when the number of transmission
blocks is sufficiently large.
\item With the sufficient condition, the roles of key system parameters
as well as design insights for such EH relay networks will be obtained.
Similar to the channel hardening effect in massive MIMO systems, a
kind of \emph{energy hardening effect} in such multi-relay EH networks
is observed. We find that when the number of EH relays is large, there
will be little performance loss if only using the instantaneous energy
side information, without the offline energy side information. 
\item We shall demonstrate via simulations that the proposed cooperation
strategy can achieve significant power savings over the direct link
transmission without EH relays. We will also show that the proposed
strategy outperforms the greedy strategy that only optimizes the current
block transmission. Specifically, the proposed strategy requires much
fewer relays to achieve the same performance, and the performance
gap between the two strategies becomes larger when the number of S-D
pairs increases. Moreover, the proposed sub-optimal algorithms provide
performance close to the corresponding performance upper bound, and
they have the potential to be extended to other cooperative EH networks. 
\end{itemize}
The rest of this paper is organized as follows. We shall describe
the system model and problem formulation in Section II. We will first
solve the respective feasibility problem with a single S-D pair in
Section III. The complete solution for the original optimization problem
is provided in Section IV. In Section V, simulation results are then
presented to demonstrate the advantage of the proposed design approach.
Finally, Section VI concludes the paper.

\emph{Notation:} Matrices are denoted by bold-face upper-case letters.
Sets are denoted by calligraphic upper-case letters such as $\mathcal{A}$,
and $\left|\mathcal{A}\right|$ represents its cardinality. \textbf{0}
denotes an all-zero matrix. $x\leftarrow y$ represents assigning
the value of $y$ to $x$. For two sets $\mathcal{A}$ and $\mathcal{B}$,
$\mathcal{A}-\mathcal{B}$ is the relative complement of $\mathcal{A}$
with respect to $\mathcal{B}$. $\mathbf{1}_{\varphi}$ denotes the
indicator function. $\varphi_{1}\Leftrightarrow\varphi_{2}$ means
that $\varphi_{1}$ is sufficient and necessary for $\varphi_{2}$.
$\left\lceil x\right\rceil $ ($\left\lfloor x\right\rfloor $ ) means
the ceiling (floor) function.

\section{System Model and Problem Formulation}

\begin{figure}
\begin{centering}
\includegraphics[scale=0.56]{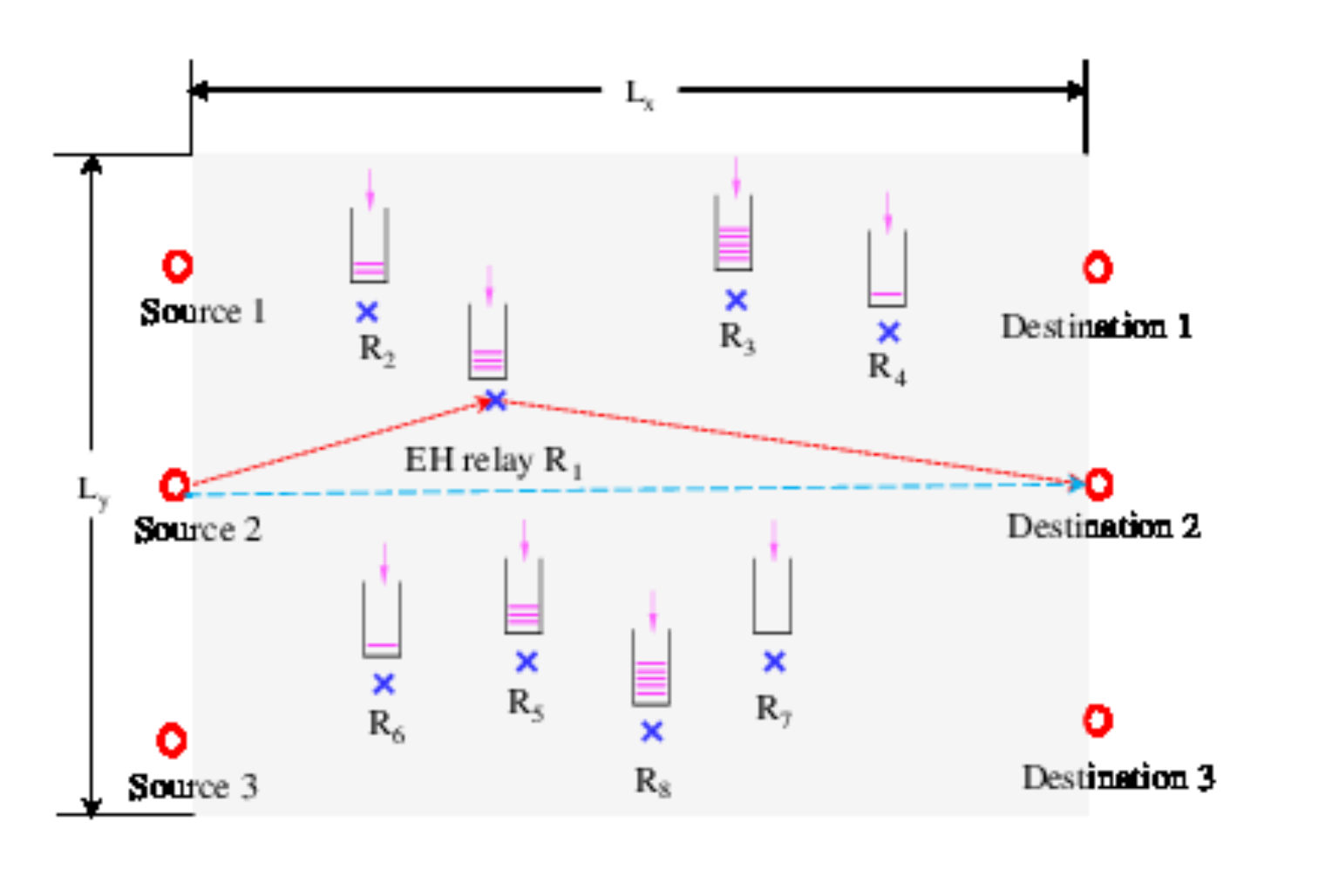}
\par\end{centering}

\protect\caption{\label{fig:system_model}An example system with 3 S-D pairs, where
the sources are non-EH nodes. To reduce the transmit power of this
network, we deploy 8 EH relays. }
\end{figure}

The main design objective of this paper is to reduce the power consumption
in an existing wireless network with $M$ source-destination (S-D)
pairs. The proposed approach is to deploy $K$ EH relays to forward
information for these S-D pairs. Consequently, the considered network
becomes a relay-assisted multi-source multi-destination network, as
shown in Fig. \ref{fig:system_model}. The set of S-D pair indices
is denoted as $\mathcal{M}=\left\{ 1,2,...,M\right\} $, and the set
of relay indices is denoted as $\mathcal{K}=\left\{ 1,2,...,K\right\} $.
In the following part of this section, we will first describe the
details of the system model, and present the proposed cooperation
strategy. To optimize the cooperation strategy, we will then formulate
the design problem to be tackled in this paper.

\subsection{System Setting}

We assume that all the channels are block fading, with the coherence
time denoted as $T^{\textrm{C}}$, corresponding to one transmission
block. In the first half of one transmission block, each source transmits
its information signal, while in the second half, the selected relay
forwards the information to the corresponding destination. We consider
the system design within a given transmission duration of length $T$,
which consists of $N=\frac{T}{T^{\textrm{C}}}$ transmission blocks,
with $N$ normally much larger than $K$. Denote the set of all the
transmission block indices in $T$ as $\mathcal{N}=\left\{ 1,2,...,N\right\} $.
To guarantee satisfactory performance of all the S-D pairs, we assume
that for each transmission block, each source needs to transmit with
the assistance of relays. Meanwhile, a certain QoS requirement is
used to regulate the communication performance. To be practical, we
assume that the source nodes do not have instantaneous channel state
information as the EH relays may not have enough energy for channel
training/feedback. Therefore, the design will be based on the statistical
channel information and the EH rate of each relay. 

Each source is assumed to use a fixed transmit power, denoted as $P_{\textrm{s},m}^{\textrm{tr}}$
for the $m$-th source, but the value of $P_{\textrm{s},m}^{\textrm{tr}}$
needs to be properly determined. Denote the transmit power vector
of all sources as $\mathbf{P}_{\textrm{s}}=\left[P_{\textrm{s},m}^{\textrm{tr}}\right]$.
In this paper, we assume that different S-D pairs are coordinated
to transmit within different frequency bands, and thus there will
be no co-channel interference \citep{Cooperative_Diversity,PA_multi_user_relay}.
Even for such an interference-free case, the design problem turns
out to be challenging, and the obtained results will be helpful for
investigating more general cases with interference. All the relays
are half-duplex and apply the amplify-and-forward (AF) relaying protocol,
while the extension to other relaying protocols is straightforward.
There is a per band peak transmit power constraint for each relay,
denoted as $P_{k,\textrm{max}}^{{\rm {tr}}}$ for the $k$-th relay
for each frequency band it uses. For reference, the main notations
are listed in Table I.

\begin{table}
\protect\caption{Main notations}

\centering{}{\small{}}%
\begin{tabular}{|c||l|}
\hline 
{\small{}Symbols} & {\small{}Definition}\tabularnewline
\hline 
\hline 
{\small{}$\mathcal{M}$} & {\small{}S-D pair set with cardinality $M$}\tabularnewline
\hline 
{\small{}$\mathcal{K}$} & {\small{}Relay set with cardinality $K$}\tabularnewline
\hline 
$\mathcal{N}$ & {\small{}Transmission block set with cardinality $N$}\tabularnewline
\hline 
{\small{}$T^{\textrm{C}}$} & {\small{}Transmission block length}\tabularnewline
\hline 
{\small{}$T^{\textrm{E}}$} & {\small{}Energy harvesting interval}\tabularnewline
\hline 
{\small{}$T$} & {\small{}Total transmission duration }\tabularnewline
\hline 
{\small{}$N$} & {\small{}Number of blocks in $T$}\tabularnewline
\hline 
{\small{}$E_{k,\Sigma}^{\textrm{EH}}\left(t\right)$} & {\small{}EH profile at time $t$ of the $k$-th relay}\tabularnewline
\hline 
{\small{}$E_{k}^{{\rm {init}}}$} & {\small{}Initial energy of the $k$-th relay}\tabularnewline
\hline 
{\small{}$P_{k,n}^{\textrm{EH}}$} & {\small{}EH rate of $k$-th relay in $n$-th transmission block}\tabularnewline
\hline 
{\small{}$E_{k,\Sigma}\left(t\right)$} & {\small{}Cumulative energy at time $t$ for the $k$-th relay}\tabularnewline
\hline 
{\small{}$P_{k,\textrm{max}}^{{\rm {tr}}}$} & {\small{}Maximum transmit power of the $k$-th relay}\tabularnewline
\hline 
\end{tabular}
\end{table}

\subsection{Energy Harvesting Model}

An important factor that determines the performance of an EH communication
node is the \textit{EH profile}, which models the cumulative harvested
energy up to time $t$. For the $k$-th EH relay, denote the EH profile
as $E_{k,\Sigma}^{\textrm{EH}}\left(t\right)$, and the initial energy
in the battery as $E_{k}^{{\rm {init}}}=E_{k,\Sigma}^{\textrm{EH}}\left(0\right)$.
We will consider a piece-wise constant model for the EH profile. That
is, each EH node has a steady but low EH rate, and the EH rate for
the $k$-th relay in the $n$-th transmission block is denoted as
$P_{k,n}^{\textrm{EH}}$. The change of the EH rate is in the time
unit of an EH interval $T^{\textrm{E}}$. As the EH rate usually does
not vary frequently, $T^{\textrm{E}}$ is normally much larger than
the channel coherence time $T^{\textrm{C}}$. A similar EH model is
adopted in \citep{Outage_min_EH_DP,EH_RS_LYM}, and it can be used
for such energy sources as solar energy \citep{Predict_EH}. We assume
$T^{\textrm{E}}=N^{\textrm{C}}T^{\textrm{C}}$, i.e. the $j$-th EH
interval includes the transmission blocks $\left(j-1\right)T^{\textrm{C}}+1$
to $jT^{\textrm{C}}$, where $j=1,2,...,N^{\textrm{E}}$. Moreover,
$T=N^{\textrm{E}}T^{\textrm{E}}$, and $N=N^{\textrm{E}}N^{\textrm{C}}$.
For simplicity, we denote the set of all the EH interval indices as
$\mathcal{J}=\left\{ 1,2,...,N^{\textrm{E}}\right\} $. As the EH
rate, which we should denote by $\Psi_{k,j}$, stays constant in the
$j$-th EH interval, then
\begin{equation}
P_{k,n_{1}}^{\textrm{EH}}=P_{k,n_{2}}^{\textrm{EH}}=\Psi_{k,j},\label{eq:1-2}
\end{equation}
$\forall n_{1},n_{2},$ such that $\left\lceil \frac{n_{1}}{N^{\textrm{C}}}\right\rceil =\left\lceil \frac{n_{2}}{N^{\textrm{C}}}\right\rceil =j$.
Moreover, we assume that the EH process for each relay is a stationary
random process. The EH rates for different relays or among different
EH intervals, i.e. $P_{k,n}^{\textrm{EH}}$ with different $k$ or
$n$, are independent of each other. Similar to \citep{EH_DWF_channel,EH_Broad_time},
we assume that offline EH information of all EH relays is available,
which is valid for some predictable energy resources, such as solar
energy. With such information, we can obtain a performance upper bound
for the case with causal energy information, and also reveal some
properties that provide design guidelines for more general cases.
In Section III.B, we find that when the number of EH relays is large
enough, the dependence on the future energy information is negligible,
which provides the possibility of only using the current energy side
information in the multi-relay EH networks. The full solution for
general EH networks with causal energy information will be left to
our future work.

For an EH node, the utilization of the harvested energy is constrained
by the EH profile, which yields the \emph{energy causality constraint}
\citep{EH_two_hop5}. The energy causality means that the energy consumed
thus far cannot exceed the total harvested energy. For the $k$-th
EH relay, denote the instantaneous power consumption as ${P_{k}}\left(t\right)$,
and then the energy causality constraint can be expressed as 
\begin{equation}
\int_{0}^{t}{P_{k}}\left(\tau\right)d\tau\le E_{k,\Sigma}^{\textrm{EH}}\left(t\right).\label{eq:1}
\end{equation}
This constraint will bring major design challenges for the EH networks.
In particular, for the system in Fig. \ref{fig:system_model}, it
will cause a coupling effect when determining the transmit powers
of different relays and selecting different relays, as will be discussed
in the following sub-section. 

In this paper, we only consider the energy consumption for information
transmission, while ignoring other types of energy consumption. For
each EH relay, we assume that the battery capacity is large enough,
while the more general case is left for future work. Moreover, we
assume that the initial energy in the battery for the $k$-th relay
can support its maximum transmit power in all the frequency bands
it supports, i.e. $E_{k}^{{\rm {init}}}=MP_{k,\textrm{max}}^{{\rm {tr}}}T^{\textrm{C}}/2$.
This is mainly to guarantee the performance during the first few transmission
blocks.

\subsection{Proposed Cooperation Strategy}

In the proposed cooperation strategy, different relays will take turns
to assist the transmissions of all the S-D pairs. The main components
of the proposed cooperation strategy include power assignment and
relay selection, which are jointly determined.

\subsubsection{Power Assignment}

As our transmit power optimization design is based on statistical
channel information, we do not consider the power adaptation of each
node over time. Once the transmit power is determined, the value will
be fixed within the whole transmission duration, and the relay will
use the same transmit power whenever it is selected to forward information.
\emph{Power assignment} determines the transmit power for each relay.
For a given relay, when assisting a certain S-D pair, a low transmit
power cannot provide a good performance. Whereas, a too high transmit
power will exhaust its energy too soon, implying that some other relays
will be selected more often. Besides this, there also exists a coupling
effect among different S-D pairs, as each relay possibly serves multiple
S-D pairs with only limited energy. The coupling effect complicates
the system design, and thus the power assignment needs to be carefully
decided. We denote the transmit power matrix of all relays as $\ensuremath{\mathbf{P}_{\textrm{r}}=\left[P_{m,k}^{{\rm {tr}}}\right]}$,
where $P_{m,k}^{\textrm{tr}}$ is the transmit power of the $k$-th
relay when assisting the $m$-th S-D pair.

\subsubsection{Relay Selection}

For the proposed cooperation strategy, relay selection will be performed
in each transmission block, and it should be based on each relay's
available energy. A relay is called \emph{active} if its available
energy, denoted as $E_{k,\Sigma}\left(t\right)$, is enough to support
its assigned transmit power, i.e. 
\begin{equation}
E_{k,\Sigma}\left(t\right)\geq P_{m,k}^{\textrm{tr}}T^{\textrm{C}}/2.\label{eq:2}
\end{equation}
In each transmission block, one active relay will be selected to assist
each S-D pair. Denote the relay selection matrix as $\boldsymbol{\textrm{Z}}=\left[z_{m,k,n}\right]$
where

\vspace{-15pt}

\[
z_{m,k,n}=\left\{ \begin{array}{cl}
1 & \textrm{\ensuremath{m}-th pair selects }k\textrm{-th relay}\textrm{ in \ensuremath{n}-th block}\\
0 & \textrm{otherwise}
\end{array}\right..
\]
Given $\boldsymbol{\textrm{Z}}$, the available energy for the $k$-th
relay before the relay transmission stage of the $l$-th transmission
block can be expressed as 
\begin{align}
 & E_{k,\Sigma}\left(\left({l-\frac{1}{2}}\right){T^{\textrm{C}}}\right)=E_{k}^{{\rm {init}}}+\sum\limits _{n=1}^{l-1}P_{k,n}^{{\rm {EH}}}{T^{\textrm{C}}}\nonumber \\
 & \:\:+P_{k,l}^{{\rm {EH}}}\frac{1}{2}{T^{\textrm{C}}}-\sum\limits _{m=1}^{M}\sum\limits _{n=1}^{l-1}{{z_{m,k,n}}P_{m,k}^{{\rm {tr}}}\frac{1}{2}{T^{\textrm{C}}}}.\label{eq:3}
\end{align}
Since ${\sum\limits _{k=1}^{K}{z_{m,k,n}}\leq1}$, $n\in\mathcal{N},$
$m\in\mathcal{M}$, we can at most select one relay for each pair
in each transmission block, and thus the decision of whether to select
one relay for a certain transmission block is coupled with that of
other relays. For each feasible relay selection matrix $\boldsymbol{\textrm{Z}}$,
different relays would take turns to forward the source information. 
\begin{rem}
With the proposed cooperation strategy, different relays will take
turns to assist the S-D communications. Consequently, each of them
will have time to accumulate enough energy for transmission, despite
the limited harvested energy in each single transmission block. This
strategy bears a similar motivation as \emph{diversity} in wireless
communications \citep{Fund_WC_David}. With diversity, multiple copies
of the same information will be sent through links with independent
fading. Hence, it is very unlikely that all the links will be weak.
Therefore, a so-called \emph{diversity gain} is achieved. On the
other hand, for our proposed cooperation strategy, the hope is that
we can always find an EH relay with enough energy to help the S-D
communication, and the achieved performance gain can be regarded as
the \emph{energy diversity gain} \citep{EH_diversity_WCSP}.
\end{rem}

\subsection{Problem Formulation}

To guarantee satisfactory performance of all the S-D pairs, we assume
that there is a QoS constraint for each pair, and it is measured by
a general utility function \citep{QoS_worst_utilityRA}, which is
a monotonically increasing function of the allocated resource. The
utility may represent the successful transmission probability, transmission
rate, or other QoS metric, depending on the application scenario.
For the $m$-th source with transmit power $P_{\textrm{s},m}^{\textrm{tr}}$,
when the $k$-th relay assists it with transmit power $P_{m,k}^{\textrm{tr}}$,
the achieved utility is denoted as $U_{m,k}\left(P_{\textrm{s},m}^{\textrm{tr}},P_{m,k}^{\textrm{tr}}\right)$.
Based on the relay selection matrix $\boldsymbol{\textrm{Z}}$, for
the $m$-th S-D pair in the $n$-th transmission block, with $m\in\mathcal{M}$,
$n\in\mathcal{N}$, we have $U^{\left(m,n\right)}=\sum_{k=1}^{K}z_{m,k,n}U_{m,k}\left(P_{\textrm{s},m}^{\textrm{tr}},P_{m,k}^{\textrm{tr}}\right).$
In order to guarantee steady and reliable communications, we assume
that the QoS constraint shall be satisfied in each transmission block
for each S-D pair. Thus, the QoS constraint is expressed as $U^{\left(m,n\right)}\ge U_{\textrm{th}}.$
Upon this, ${\sum\limits _{k=1}^{K}{z_{m,k,n}}=1}$ shall hold for
all $n\in\mathcal{N},$ $m\in\mathcal{M}$, i.e. each pair will select
one relay in each transmission block. Otherwise, the QoS requirement
will be violated.

The design objective is to minimize the maximum transmit power among
all the S-D pairs, i.e. to attain power conservation for all the sources
at the same time. The design problem can then be formulated as 

\vspace{-15pt}

\begin{align}
\textrm{\textbf{OP1:}}\:\nonumber \\
{\mathop{\min}\limits _{\mathbf{P}_{\textrm{s}},\,\mathbf{P}_{\textrm{r}},\,\mathbf{Z}}} & \phantom{=}\mathop{\max}\limits _{m}\:{P_{m,s}^{{\rm {tr}}}}\nonumber \\
\textrm{s.t.} & \phantom{=}{\sum\limits _{m=1}^{M}\sum\limits _{n=1}^{l}{{z_{m,k,n}}P_{m,k}^{{\rm {tr}}}\frac{1}{2}{T^{\textrm{C}}}}\le E_{k}^{{\rm {init}}}+\sum\limits _{n=1}^{l-1}P_{k,n}^{{\rm {EH}}}{T^{\textrm{C}}}}\nonumber \\
 & \phantom{=}\:\:+P_{k,l}^{{\rm {EH}}}\frac{1}{2}{T^{\textrm{C}}},\:\forall l\in\mathcal{N},\: k\in\mathcal{K},\label{eq:6}\\
 & \phantom{=}{P_{m,k}^{\textrm{tr}}\le P_{k,\textrm{max}}^{{\rm {tr}}},\:\forall k\in\mathcal{K},\: m\in\mathcal{M},}\label{eq:7}\\
 & \phantom{=}{U^{\left(m,n\right)}\ge U_{\textrm{th}},\:\forall n\in\mathcal{N},\: m\in\mathcal{M},}\label{eq:8}\\
 & \phantom{=}{{\sum\limits _{k=1}^{K}{z_{m,k,n}}=1},\:\forall n\in\mathcal{N},\: m\in\mathcal{M},}\label{eq:9}\\
 & \phantom{=}z_{m,k,n}\in\left\{ 0,1\right\} ,\:\forall m\in\mathcal{M},\: k\in\mathcal{K},\: n\in\mathcal{N},\label{eq:10}
\end{align}
where (\ref{eq:6}) is based on the energy causality constraint (\ref{eq:1}),
(\ref{eq:7}) is the peak power constraint for each relay, (\ref{eq:8})
is the QoS constraint, and (\ref{eq:9}) guarantees that for each
transmission block, one relay is selected for each S-D pair. 

\textbf{OP1} is a joint power assignment and relay selection optimization
problem, i.e. we jointly design $\mathbf{P}$ and $\mathbf{Z}$. It
is a highly complicated problem, as it belongs to the mixed-integer
nonlinear programming (MINLP) problem \citep{MINLP}, which is known
to be NP-hard. Moreover, one particular difficulty is the coupling
effect among the operations for different relays, and another comes
from the large variable size. The worst-case complexity is exponential
with the source number $M$, the relay number $K$, and the transmission
block number $N$. We will consider the following equivalent epigraph
form of \textbf{OP1}

\vspace{-15pt}

\begin{align*}
\textrm{\textrm{\textbf{OP2:}}}\\
{\mathop{\min}\limits _{\mathbf{P}_{\textrm{s}},\,\mathbf{P}_{\textrm{r}},\,\mathbf{Z},\,\eta}} & \phantom{=}\eta\\
\textrm{s.t.} & \phantom{=}{{P_{m,s}^{{\rm {tr}}}\le\eta},\:\forall m\in\mathcal{M},}\\
 & \phantom{=}{\textrm{Constraints }\eqref{eq:6}\sim\eqref{eq:10}.}
\end{align*}
In the following, we will first investigate the feasibility problem
for \textbf{OP2} with a given $\eta$,

\vspace{-15pt}

\begin{align*}
\textrm{\textrm{\textbf{FP1:}}}\\
{\textrm{find}} & \phantom{=}\mathbf{P}_{\textrm{s}},\,\mathbf{P}_{\textrm{r}},\,\mathbf{Z},\\
\textrm{s.t.} & \phantom{=}{{P_{m,s}^{{\rm {tr}}}\le\eta},\:\forall m\in\mathcal{M},}\\
 & \phantom{=}{\textrm{Constraints }\eqref{eq:6}\sim\eqref{eq:10},}
\end{align*}
which will then help develop a general design framework and efficient
algorithms for the original problem. Particularly, it can be checked
that the feasibility of \textbf{FP1} with various $\eta$ has a \emph{two-phase
pattern}. If \textbf{FP1} is feasible for a given $\eta$, it must
be feasible for any $\eta'>\eta$. On the other hand, if it is infeasible
for a given $\eta$, then it must be infeasible for any $\eta'<\eta$.
Hence, we can deduce that the smallest $\eta$ that can make \textbf{FP1}
feasible corresponds to the optimal solution of the original optimization
problem \textbf{OP1}. Therefore, the major task now lies in solving
the feasibility problem \textbf{FP1} with a given $\eta$. In the
following, we will start by investigating \textbf{FP1} for the single-pair
case, i.e. $M=1$, in section III. The solution is then generalized
to solve the feasibility problem \textbf{FP1} with multiple pairs.
Based on this feasibility study, the solution to the original problem
\textbf{OP1} will be obtained. Details of the proposed algorithm for
\textbf{OP1} will be provided in section IV.

\section{A Special Case: Feasibility Study for a Single S-D Pair}

In this section, we will deal with the feasibility problem \textbf{FP1}
with a single S-D pair and a given $\eta$. For the notations in this
section, the pair index $m$ will be removed. Particularly, the relay
transmit power assignment matrix $\mathbf{P}_{\textrm{r}}$ is reduced
to a vector, and the source transmit power vector to a scalar. 

With a single S-D pair, the feasibility problem \textbf{FP1} with
a given $\eta$ is 

\vspace{-15pt}

\begin{align}
\textrm{\textrm{\textbf{FP2:}}}\nonumber \\
\textrm{find} & \phantom{=}P_{s}^{{\rm {tr}}},\,\mathbf{P}_{\textrm{r}},\,\mathbf{Z}\nonumber \\
\textrm{s.t.} & \phantom{=}{{P_{s}^{{\rm {tr}}}\le\eta},}\label{eq:11}\\
 & \phantom{=}{\sum\limits _{n=1}^{l}{{z_{k,n}}P_{k}^{{\rm {tr}}}\frac{1}{2}{T^{\textrm{C}}}}\le E_{k}^{{\rm {init}}}+\sum\limits _{n=1}^{l-1}P_{k,n}^{{\rm {EH}}}{T^{\textrm{C}}}}\nonumber \\
 & \phantom{=}\:\:+P_{k,l}^{{\rm {EH}}}\frac{1}{2}{T^{\textrm{C}}},\:\forall l\in\mathcal{N},\: k\in\mathcal{K},\label{eq:12}\\
 & \phantom{=}{P_{k}^{\textrm{tr}}\le P_{k,\textrm{max}}^{{\rm {tr}}},\:\forall k\in\mathcal{K},}\label{eq:13}\\
 & \phantom{=}{U^{\left(n\right)}\ge U_{\textrm{th}},\:\forall n\in\mathcal{N},}\label{eq:13-1}\\
 & \phantom{=}{{\sum\limits _{k=1}^{K}{z_{k,n}}=1},\:\forall n\in\mathcal{N},}\label{eq:14}\\
 & \phantom{=}{z_{k,n}\in\left\{ 0,1\right\} ,\:\forall k\in\mathcal{K},\: n\in\mathcal{N}.}\nonumber 
\end{align}
In the following, we will first reformulate this problem into a simpler
form, which is still NP-hard, but can help derive a sufficient condition
for its feasibility. The scenarios where the sufficient condition
becomes necessary will also be tackled.

\subsection{Problem Reformulation}

In this sub-section, we will remove constraints (\ref{eq:11}) and
(\ref{eq:13-1}) by introducing the operation of relay pre-selection
and fixing the relay transmit power $\mathbf{\mathbf{P}_{\textrm{r}}}$,
respectively. Meanwhile, the variable size will be greatly reduced.

With a given $\eta$, for the $k$-th relay, if ${U_{k}\left(\eta,P_{k,\textrm{max}}^{\textrm{tr}}\right)}<U_{\textrm{th}}$,
then this relay should not be selected even once, i.e. $z_{k,n}=0,$
$\forall n\in\mathcal{N}$. Otherwise, constraint (\ref{eq:13-1})
will be violated. On the other hand, if $U_{k}\left(\eta,P_{k,\textrm{max}}^{\textrm{tr}}\right)\geq U_{\textrm{th}}$,
it will be selected at least once within the transmission duration
$T$, i.e. $\sum_{n=1}^{N}z_{k,n}\geq1$, given that $N\geq K$. Hence,
without loss of generality, we can just let the source select these
relays one by one in the first several transmission blocks. This can
be done until all these relays have been selected once, without affecting
the feasibility. 

Based on the above insights, we define the set of such relays as the
\emph{candidate relay subset}. The procedure of determining the candidate
relay subset is called \emph{relay pre-selection}. Therefore, the
temporary%
\footnote{This relay pre-selection result $\mathcal{S}_{\eta}$ is only a ``temporary''
result, as it is for a given $\eta$, while in general it will be
different from the final result.%
} relay pre-selection result with a given $\eta$ is determined as

\vspace{-10pt}
\begin{equation}
\mathcal{S}_{\eta}=\left\{ k\in\mathcal{K}\left|U_{k}\left(\eta,P_{k,\textrm{max}}^{\textrm{tr}}\right)\geq U_{\textrm{th}}\right.\right\} .\label{eq:15}
\end{equation}
For notational simplicity and with a given $\eta$ and an arbitrary
$x$, denote $g_{k,\eta}\left(x\right)=U_{k}\left(\eta,x\right)$,
and the inverse function of $g_{k,\eta}\left(x\right)$ over $x$
as $g_{k,\eta}^{-1}\left(\cdotp\right)$. For $k\in\mathcal{S}_{\eta}$,
the minimum transmit power of the $k$-th relay that can meet the
QoS constraint $U_{\textrm{th}}$ is determined as $g_{k,\eta}^{-1}\left(U_{\textrm{th}}\right)$.
For a given $\eta$, this minimum transmit power only depends on the
relay location. We denote $\hat{\mathbf{P}}_{\eta}=\left[\hat{P}_{k}\left(\eta\right)\right]$
where

\vspace{-10pt}

\begin{equation}
\hat{P}_{k}\left(\eta\right)=\begin{cases}
g_{k,\eta}^{-1}\left(U_{\textrm{th}}\right) & k\in\mathcal{S}_{\eta}\\
0 & k\in\mathcal{K}-\mathcal{S}_{\eta}
\end{cases}.\label{eq:16}
\end{equation}
Then, we have the following property that can simplify the feasibility
checking of \textbf{FP2}: 
\begin{lem}
\label{lem: Fix_Power}\textbf{FP2} is feasible $\Leftrightarrow$
\textbf{FP2} with \textup{$\mathbf{P}_{\textrm{r}}=\hat{\mathbf{P}}_{\eta}$
}is feasible.\end{lem}
\begin{IEEEproof}
We prove the equivalence in two steps. We will first show the necessity,
and then the sufficiency. The necessity is obvious, as by fixing the
transmit power, the feasible domain shrinks. If the shrunken domain
is still feasible, so is the original one. For the sufficiency part,
and based on the definition of $P_{k}^{*}\left(\eta\right)$ in (\ref{eq:16}),
$P_{k}^{*}\left(\eta\right)$ is the smallest power value satisfying
(\ref{eq:11}). With any given $\mathbf{Z}$, if $P_{k}^{*}\left(\eta\right)$
violates (\ref{eq:12}) or (\ref{eq:13}), no other power value can
satisfy (\ref{eq:12}) and (\ref{eq:13}). Therefore, if the feasible
domain with fixed $P_{k}^{\textrm{tr}}=P_{k}^{*}\left(\eta\right)$
is empty, so is the feasible domain with other power values.
\end{IEEEproof}
With Lemma \ref{lem: Fix_Power}, we can replace the power vector
$\mathbf{P}_{\textrm{r}}$ in \textbf{FP2} by $\hat{\mathbf{P}}_{\eta}$
and also remove Eq. (\ref{eq:13-1}). Therefore, when checking the
feasibility of \textbf{FP2}, we do not need to check the full domain
of the power vector $\mathbf{P}_{\textrm{r}}$, but need only to check
a single vector $\hat{\mathbf{P}}_{\eta}$, which largely reduces
the complexity. Moreover, with a given $\mathcal{S}_{\eta}$,\textbf{
}and based on the expression of the utility function $U^{\left(m,n\right)}$
in the last section, we only need to have $U_{m,k}\left(\eta,P_{k,\textrm{max}}^{\textrm{tr}}\right)\geq U_{\textrm{th}}$,
and ${\sum\limits _{k\in\mathcal{S}_{\eta}}{z_{k,n}}=1},\:\forall n\in\mathcal{N}$,
which means that there are active relays for each transmission block
and all of these relays can satisfy the QoS constraint. Therefore,
(\ref{eq:11}) can also be removed from \textbf{FP2}.

Based on the above discussion, we have fixed $\mathbf{P}_{\textrm{r}}$
and removed constraints (\ref{eq:11}) and (\ref{eq:13-1}), and thus
we obtain the following equivalent problem for \textbf{FP2}:

\vspace{-15pt}

\begin{align}
\textrm{\textrm{\textbf{FP3:}}}\nonumber \\
\textrm{\textrm{find}} & \phantom{=}\mathbf{Z}\nonumber \\
\textrm{s.t.} & \phantom{=}{\sum\limits _{n=1}^{l}{\frac{1}{2}{\hat{P}_{k}\left(\eta\right)T^{\textrm{C}}}z_{k,n}}\le E_{k}^{{\rm {init}}}+\sum\limits _{n=1}^{l-1}P_{k,n}^{{\rm {EH}}}{T^{\textrm{C}}}}\nonumber \\
 & \phantom{=}\:\:+P_{k,l}^{{\rm {EH}}}\frac{1}{2}{T^{\textrm{C}}},\:\forall l\in\mathcal{N},\: k\in\mathcal{S}_{\eta},\label{eq:17}\\
 & \phantom{=}{{\sum\limits _{k\in\mathcal{S}_{\eta}}{z_{k,n}}=1},\:\forall n\in\mathcal{N},}\: z_{k,n}\in\left\{ 0,1\right\} ,\nonumber \\
 & \phantom{=}\:\:\forall k\in\mathcal{S}_{\eta},\: n\in\mathcal{N}.\label{eq:18}
\end{align}
Compared with \textbf{FP2}, the dimension of \textbf{FP3} is reduced
as $\mathbf{P}_{\textrm{r}}$ is fixed, and some constraints are removed.
It can be verified that \textbf{FP3} is the respective feasibility
problem of a multi-resource generalized assignment problem (MRGAP)
\citep{AP_anniversary}, which is still NP-hard. A particular difficulty
is the large size of $\mathbf{Z}$. Therefore, although the problem
has been significantly simplified, it is still difficult to deal with.
In the following, by exploiting the relationship of several key system
parameters, we will derive a sufficient condition for the feasibility
of \textbf{FP3}, which is easy to check. This will then be used to
provide a sub-optimal algorithm for the original optimization problem
\textbf{OP2}.

\subsection{Condition for Feasibility}

In this part, despite the complexity of the problem, we will propose
a simple condition to check the feasibility of the problem, which
is sufficient, and will become necessary for some asymptotic scenarios. 

Define the cumulative average EH rate up to the end of the $j$-th
EH interval for the $k$-th relay as $\bar{\Psi}_{k,j}=\frac{\sum_{i=1}^{j}\Psi_{k,j}}{j}$,
$\forall j\in\mathcal{J}$. With the  given $\eta$ and $l$, denote

\vspace{-15pt}

\begin{equation}
\zeta\left(\eta,\: l\right)=\underset{\zeta_{1}\left(\eta,\: l\right)}{\underbrace{{\sum_{k\in{{\cal S}_{\eta}}}}\frac{{2\bar{\Psi}_{k,l}}}{{{{\hat{P}}_{k}}\left(\eta\right)}}}}+\underset{\zeta_{2}\left(\eta,\: l\right)}{\underbrace{{\sum_{k\in{{\cal S}_{\eta}}}}\frac{{2E_{k}^{{\rm {init}}}/T^{\textrm{C}}-{{\hat{P}}_{k}}\left(\eta\right)}}{{{{\hat{P}}_{k}}\left(\eta\right)lN^{\textrm{C}}}}}}.\label{eq:22-3}
\end{equation}
We briefly introduce the insights from the expression of $\zeta\left(\eta,\: l\right)$
in Eq. (\ref{eq:22-3}).\textbf{ }The two parts of $\zeta\left(\eta,\: l\right)$
play different roles in the system operation. The first term, $\zeta_{1}\left(\eta,\: l\right)$,
can be regarded as the steady-state component, which manipulates the
long-term%
\footnote{Here \textquotedbl{}long term\textquotedbl{} refers to the part that
still remains after an extremely large number of transmission blocks. %
} balance of the harvested energy and consumed energy. On the other
hand, the second term, $\zeta_{2}\left(\eta,\: l\right)$, is specifically
designed to improve the performance at the beginning stage of the
whole transmission duration. The value of $\zeta_{2}\left(\eta,\: l\right)$
is computed based on the approximation that all relays' harvested
energy and initial energy are exactly used up in the end. Meanwhile
the initial energy for each relay is enough for the transmit energy
of one packet. With or without the second term $\zeta_{2}\left(\eta,\: l\right)$,
the sufficiency always holds. But the performance without this term
will be poor in the case where the transmission block number is not
too large. Based on Eq. (\ref{eq:22-3}) we have the following result,
which provides a simple sufficient condition to check the feasibility
of \textbf{FP3}. 
\begin{thm}
\label{theo: Feasibility condition}The following is a sufficient
condition for the feasibility of \textbf{FP3}:

\emph{
\begin{equation}
\underset{j\in\mathcal{J}}{\textrm{min}}\:\zeta\left(\eta,\: j\right)\ge1.\label{eq:20}
\end{equation}
}Particularly, for two asymptotic scenarios, with other parameters
fixed, we have the following properties:

1. The necessity property: When \textup{$N^{\textrm{C}}\rightarrow\infty$},
the sufficient condition is necessary.

2. The deterministic property: When $K\rightarrow\infty$, \textup{$\zeta\left(\eta,\: l\right)$}
becomes deterministic%
\footnote{Here we only consider the non-trivial cases. For some $\eta$, it
is possible that $\zeta\left(\eta,\: j\right)$ becomes infinity.
But in this case, the condition is already satisfied trivially, and
the problem \textbf{FP3} can be easily verified to be feasible. This
corresponds to the case where the system is with enough harvested
energy from these relays. In this case, the value of $\eta$ can be
further reduced given the relays' EH situations, and this $\eta$
cannot be optimal for the original \textbf{OP2}. Here we ignore such
trivial cases.%
}, independent of the concrete realization of $\left[P_{k}^{{\rm {EH}}}\right]$. \end{thm}
\begin{IEEEproof}
The proof is given in the Appendix.
\end{IEEEproof}
As shown in the proof, once condition (\ref{eq:20}) is satisfied,
there exist active relays for each transmission block. In this case,
different relays will take turns to assist the S-D pair and the unsatisfactory
performance caused by the low EH rate at a single relay can be overcome.
Moreover, based on Theorem \ref{theo: Feasibility condition}, and
given that the effect of the large-sized matrix $\mathbf{Z}$ has
been removed, condition (\ref{eq:20}) only depends on $\hat{\mathbf{P}}_{\eta}$,
and thus it is easy to check.

Next, we will discuss the insights that can be obtained from Theorem
\ref{theo: Feasibility condition}:

\textbf{1. Insights from the condition (\ref{eq:20}) on the whole.}
Based on the expression of $\zeta\left(\eta,\: l\right)$ for each
given $l$, the minimization of $\zeta\left(\eta,\: l\right)$ over
$l$ in condition (\ref{eq:20}) is to adapt to different EH rates
over different EH intervals. For condition (\ref{eq:20}), we can
find that increasing $\eta$ has two effects --$\hat{P}_{k}\left(\eta\right)$
will increase according to (\ref{eq:16}), while the set $\mathcal{S}_{\eta}$
may shrink. Therefore, a larger $\eta$ makes it easier to violate
condition (\ref{eq:20}). That is, the achievable performance for
\textbf{OP2} will be limited. In order to improve it, it is better
to have more relays with higher average EH rates and lower $\hat{P}_{k}\left(\eta\right)$
(corresponding to a more favorable location).

\textbf{2. Insights from the necessity property.} When the number
of transmission blocks is large, the condition becomes sufficient
and necessary, and the performance loss caused by using this condition
instead of directly solving \textbf{FP3} disappears. Moreover, the
later simulation results will show that the convergence speed is very
fast. In particular, to approach the close-to-optimal performance,
only tens of transmission blocks are enough, which can be naturally
satisfied by the practical system parameters.

\textbf{3. Insights from the deterministic property.} When the relay
number is large, besides the energy diversity mentioned in the last
section, another phenomenon occurs. That is, the randomness from each
single EH relay vanishes gradually. Similar to the \emph{channel hardening}
effect in massive MIMO systems \citep{Channel_hardening}, this kind
of effect can be called  the \emph{energy hardening} effect. The offline
EH information, which is very important for the optimal design in
EH networks \citep{EH_DWF_time}, is usually not easy to obtain in
practical systems, especially when the number of EH nodes is large.
Fortunately, the energy hardening effect makes it possible that the
performance loss due to the lack of offline EH information becomes
extremely small, as long as the relay number is large. This will avoid
the necessity of predicting future EH rates for systems with a large
number of relays, and thus will greatly simplify the protocol designs.

\section{Proposed Algorithm for Transmit Power Minimization}

In this section, we will first investigate the feasibility problem
with multiple S-D pairs, with the help of the results for the single-pair
case. We will then propose an efficient bisection algorithm to solve
the original joint power assignment and relay selection problem.

\subsection{The Feasibility Problem with Multiple S-D pairs}

In this subsection, we will investigate the feasibility problem with
multiple S-D pairs. Compared with the single S-D pair case, a special
difficulty lies in the coupling effect among different S-D pairs as
each relay is expected to assist multiple pairs during the whole transmission
duration. At first, we will divide the problem into several sub-problems
by introducing new variables. This procedure can be physically understood
as relay division. Next, for each sub-problem, we exploit the results
for the single-pair case to simplify it. Finally, jointly considering
all the sub-problems together, the complexity is further squeezed.
Note that, among the previously-mentioned steps, all the other steps
preserve the equivalence except the one applying the single-pair case
result, which contributes only sub-optimality.

\subsubsection{Problem Decoupling}

In this part, we will reformulate the problem into a decoupled version
by introducing new variables, and this procedure will be later physically
interpreted as a procedure of relay division. 

We are faced with a relay set in which an arbitrary relay is possible
to serve an arbitrary S-D pair, and each relay needs to determine
its power allocation for different pairs. Therefore, for a certain
relay, the decision of how to assist one S-D pair is coupled with
the decision for another pair. Such a coupling effect greatly complicates
the problem. Note that the coupling effect among different S-D pairs
is caused by constraint (\ref{eq:6}). Fortunately, by introducing
two new variable sets $\left[\phi_{m,k}\right]$ and $\left[\theta_{m,k,n}\right]$,
it can be verified that the feasibility of (\ref{eq:6})\textbf{ }is
equivalent to that of the following decoupled one.

\vspace{-15pt}

\begin{equation}
\begin{array}{l}
\sum\limits _{n=1}^{l}{{z_{m,k,n}}P_{m,k}^{{\rm {tr}}}\frac{{T^{\textrm{C}}}}{2}}\le{\phi_{m,k}}E_{k}^{\textrm{init}}+\sum\limits _{n=1}^{l-1}{{\theta_{m,k,n}}P_{k,n}^{{\rm {EH}}}{T^{\textrm{C}}}}\\
\:\:\:+{\theta_{m,k,l}}P_{k,l}^{{\rm {EH}}}\frac{{T^{\textrm{C}}}}{2},\:\forall l,\: k,\: m,\\
\phantom{=}0\leq\phi_{m,k}\leq1,\:0\leq{\theta_{m,k,n}}\leq1,\:\forall m\in\mathcal{M},\: k\in\mathcal{K},\: n\in\mathcal{N},\\
\phantom{=}{\sum\limits _{m=1}^{M}{\phi_{m,k}}=1,}\:{\sum\limits _{m=1}^{M}{\theta_{m,k,n}}=1,\:\forall k\in\mathcal{K},\: n\in\mathcal{N}.}
\end{array}\label{eq:23-1}
\end{equation}

\begin{rem}
The previous transformation from (\ref{eq:6}) to (\ref{eq:23-1})
has a clear physical meaning as relay division. We consider the energy
flow through a certain relay. Based on Eq. (\ref{eq:6}), for the
$k$-th relay, the initial energy is $E_{k}^{\textrm{init}}$. In
the $n$-th transmission block, the entering energy is from the EH
process with EH rate $P_{k,n}^{{\rm {EH}}}$, while the exiting part
includes the transmit powers for all the S-D pairs it possibly serves,
i.e. $\sum\nolimits _{m=1}^{M}{{z_{m,k,n}}P_{m,k}^{{\rm {tr}}}}$.
While for Eq. (\ref{eq:23-1}), with given values of $\left[\phi_{m,k}\right]$
and $\left[\theta_{m,k,n}\right]$, we can equivalently interpret
(\ref{eq:23-1}) as an individual relay only serving the $m$-th S-D
pair, and is denoted as the $\left(m,k\right)$-th \emph{child relay},
with initial energy as ${\phi_{m,k}}E_{k}^{\textrm{init}}$, harvested
power as ${\theta_{m,k,n}}P_{k,n}^{{\rm {EH}}}$, and transmit power
as $P_{m,k}^{{\rm {tr}}}$. To avoid any confusion, the original relay
is called the \emph{parent relay}. In this case, through $\left[\phi_{m,k}\right]$
and $\left[\theta_{m,k,n}\right]$, the original relay set $\mathcal{K}$
is equivalently transformed into a new set including only the $MK$
child relays. 
\end{rem}
Based on the previous transformation, with given $\left[\phi_{m,k}\right]$
and $\left[\theta_{m,k,n}\right]$, the original feasibility problem
\textbf{FP1} can be divided into $M$ sub-problems, and the $m$-th
($m\in\mathcal{M}$) sub-problem is

\vspace{-15pt}

\begin{align*}
\textrm{\textrm{\textbf{FP4:}}}\\
\textrm{find} & \phantom{=}P_{m,s}^{{\rm {tr}}},\, P_{m,k}^{{\rm {tr}}},\, z_{m,k,n},\,\forall k\in\mathcal{K},\: n\in\mathcal{N}\\
\textrm{s.t.} & \phantom{=}{{P_{m,s}^{{\rm {tr}}}\le\eta},}\\
 & \phantom{=}{\sum\limits _{n=1}^{l}{{z_{m,k,n}}P_{m,k}^{{\rm {tr}}}\frac{{T^{\textrm{C}}}}{2}}\le{\phi_{m,k}}E_{k}^{\textrm{init}}+\sum\limits _{n=1}^{l-1}{{\theta_{m,k,n}}P_{k,n}^{{\rm {EH}}}{T^{\textrm{C}}}}}\\
 & \phantom{=}\:\:+{\theta_{m,k,l}}P_{k,l}^{{\rm {EH}}}\frac{{T^{\textrm{C}}}}{2},\:\forall l\in\mathcal{N},\: k\in\mathcal{K},\\
 & \phantom{=}{P_{m,k}^{\textrm{tr}}\le P_{k,\textrm{max}}^{{\rm {tr}}},\:{U_{m,k}\left(P_{\textrm{s},m}^{\textrm{tr}},P_{m,k}^{\textrm{tr}}\right)}\ge U_{\textrm{th}},\:\forall k\in\mathcal{K},}\:\\
 & \phantom{=}{{\sum\limits _{k=1}^{K}{z_{m,k,n}}=1},\:\forall n\in\mathcal{N},}\:\\
 & \phantom{=}z_{m,k,n}\in\left\{ 0,1\right\} ,\:\forall k\in\mathcal{K},\: n\in\mathcal{N}.
\end{align*}

\subsubsection{Exploiting the Solution for Single S-D Pair}

For the $m$-th sub-problem in \textbf{FP4}, only a single S-D pair
is involved, and we can use the results of Theorem \ref{theo: Feasibility condition}.
Similarly, we have the temporary candidate relay subset for the $m$-th
S-D pair:

\vspace{-10pt}

\begin{equation}
{\cal S}_{\eta,m}=\left\{ k\in\mathcal{K}\left|U_{m,k}\left(\eta,P_{k,\textrm{max}}^{\textrm{tr}}\right)\geq U_{\textrm{th}}\right.\right\} ,\label{eq:29-2}
\end{equation}
and the cumulative average EH rate of the $\left(m,k\right)$-th child
relay up to the $j$-th EH interval as

\vspace{-10pt}
\begin{equation}
\bar{P}_{m,k,j}^{{\rm {EH}}}=\frac{\sum_{i=1}^{jN^{\textrm{C}}}{\theta_{m,k,i}}P_{k,i}^{\textrm{EH}}}{jN^{\textrm{C}}}.\label{eq:29-1}
\end{equation}

Moreover, with a given $\eta$ and an arbitrary $x$, denote $g_{m,k,\eta}\left(x\right)=U_{m,k}\left(\eta,x\right)$,
and the inverse function $g_{m,k,\eta}\left(x\right)$ over $x$ as
$g_{m,k,\eta}^{-1}\left(\cdotp\right)$. For $k\in\mathcal{S}_{\eta,m}$,
the minimum transmit power of the $k$-th relay that can meet the
QoS requirement $U_{\textrm{th}}$ for the $m$-th S-D pair is determined
as $g_{m,k,\eta}^{-1}\left(U_{\textrm{th}}\right)$. We denote $\hat{\mathbf{P}}_{\eta}=\left[\hat{P}_{m,k}\left(\eta\right)\right]$
where

\vspace{-10pt}
\begin{equation}
\hat{P}_{m,k}\left(\eta\right)=\begin{cases}
g_{m,k,\eta}^{-1}\left(U_{\textrm{th}}\right) & k\in{\cal S}_{\eta,m}\\
0 & k\in\mathcal{K}-{\cal S}_{\eta,m}
\end{cases}.\label{eq:16-1}
\end{equation}

With the given $\eta$ and $l$, denote $\zeta_{m}\left(\eta,\: l\right)={\sum_{k\in{{\cal S}_{\eta,m}}}}\frac{{2\bar{P}_{m,k,l}lN^{\textrm{C}}+2{\phi_{m,k}}E_{k}^{{\rm {init}}}/T^{\textrm{C}}-2{{\hat{P}}_{k}}\left(\eta\right)}}{{{{\hat{P}}_{m,k}}\left(\eta\right)lN^{\textrm{C}}}}$.
Referring to Theorem \ref{theo: Feasibility condition}, we can verify
that the feasibility of the following problem is necessary for that
of \textbf{FP4}, jointly considering all the $M$ sub-problems in
\textbf{FP4}: 

\begin{align*}
\textrm{\textrm{\textbf{FP5:}}} &  & \phantom{={2\bar{P}_{m,k,j}^{{\rm {EH}}}jN^{\textrm{C}}+2{\phi_{m,k}}E_{k}^{\textrm{init}}/T^{\textrm{C}}-2{{\hat{P}}_{k}}\left(\eta\right)}\mathcal{M}}
\end{align*}

\vspace{-15pt}

\begin{align}
\textrm{find} & \phantom{=}\phi_{m,k},\,\theta_{m,k,n},\,\forall k\in\mathcal{K},\: n\in\mathcal{N},\: m\in\mathcal{M}\nonumber \\
\textrm{s.t.} & \phantom{=}{\sum_{k\in{{\cal S}_{\eta,m}}}}\frac{{2\bar{P}_{m,k,j}^{{\rm {EH}}}jN^{\textrm{C}}+2{\phi_{m,k}}E_{k}^{\textrm{init}}/T^{\textrm{C}}-2{{\hat{P}}_{k}}\left(\eta\right)}}{{jN^{\textrm{C}}{{\hat{P}}_{m,k}}\left(\eta\right)}}\ge1,\nonumber \\
 & \phantom{=}\:\:\forall j\in\mathcal{J},\: m\in\mathcal{M},\label{eq:28-1}\\
 & \phantom{=}0\leq\phi_{m,k}\leq1,\:0\leq{\theta_{m,k,n}}\leq1,\:\forall m\in\mathcal{M},\nonumber \\
 & \phantom{=}\:\: k\in\mathcal{K},\: n\in\mathcal{N},\label{eq:7-1-1}\\
 & \phantom{=}{\sum\limits _{m=1}^{M}{\phi_{m,k}}=1,}\:{\sum\limits _{m=1}^{M}{\theta_{m,k,n}}=1,}\nonumber \\
 & \phantom{=}\:\:\forall k\in\mathcal{K},\: n\in\mathcal{N}.\label{eq:35}
\end{align}
Complexity is greatly squeezed from \textbf{FP1} to \textbf{FP5},
as the complexity for\textbf{ FP5} is linear with $\sqrt{N}MK$\textbf{.
}However, \textbf{FP5} still suffers from the large size of those
new variables $\left[\phi_{m,k}\right]$ and $\left[\theta_{m,k,n}\right]$.
Next we will try to further reduce this complexity by investigating
the special properties of the problem.

\subsubsection{Complexity Reduction}

Note that $\left[\theta_{m,k,n}\right]$ is a three-dimensional matrix,
and the size of the third dimension $n$ is possible to be extremely
large in practice. Whereas, we find that its size can be greatly reduced
with the piece-wise constant property of the EH profile. The main
result is listed in the following lemma. 
\begin{lem}
\label{lem:Fix theta}The feasibility of \textbf{FP5 }is uninfluenced
by setting constant values for $\theta_{m,k,n}$ within one EH interval,
i.e. setting $\theta_{m,k,n_{1}}=\theta_{m,k,n_{2}}=\theta_{m,k}^{j},$
$\forall n_{1},n_{2},$ such that \textup{$\left\lceil \frac{n_{1}}{N^{\textrm{C}}}\right\rceil =\left\lceil \frac{n_{2}}{N^{\textrm{C}}}\right\rceil =j$},
\textup{$1\leq j\leq N^{\textrm{E}}$}. Thus, we can replace (\ref{eq:29-1})
with the following expression:

\vspace{-15pt}

\begin{equation}
\bar{P}_{m,k,j}^{{\rm {EH}}}=\frac{\sum_{i=1}^{j}{\theta_{m,k}^{i}}\Psi_{k,i}}{j}.\label{eq:28-2}
\end{equation}
\end{lem}
\begin{IEEEproof}
Because $\forall n_{1},n_{2},$ such that $\left\lceil \frac{n_{1}}{N^{\textrm{C}}}\right\rceil =\left\lceil \frac{n_{2}}{N^{\textrm{C}}}\right\rceil =j$,
$P_{k,n_{1}}^{\textrm{EH}}=P_{k,n_{2}}^{\textrm{EH}}=\Psi_{k,j}.$
Thus, for both (\ref{eq:29-1}) and (\ref{eq:20}) only the sum of
such $\theta_{m,k,n}$ within each EH interval matters, i.e. $\sum_{n=lN^{\textrm{C}}+1}^{\left(l+1\right)N^{\textrm{C}}}\theta_{m,k,n}$
for each $l$ such that $1\leq l\leq N^{\textrm{E}}$. 
\end{IEEEproof}
Using Lemma \ref{lem:Fix theta}, matrix $\left[\theta_{m,k,n}\right]$
with size $M\times K\times N$ is greatly reduced to $\left[\theta_{m,k}^{j}\right]$
with size $M\times K\times N^{\textrm{E}}$. However, the complexity
is still high. In particular, the feasibility of Eq. (\ref{eq:28-2})
for a certain $j=j^{*}\leq N^{\textrm{E}}$ is largely influenced
by all the values of $\theta_{m,k}^{j}$ for $1\leq j\leq j^{*}$.
That is, the decisions of $\left[\theta_{m,k}^{j}\right]$ for different
EH intervals (i.e. different $j$) are coupled with each other. To
resolve this issue, we will propose a method to decouple $\left[\theta_{m,k}^{j}\right]$
at different EH intervals. The main results are listed in the following
lemma.
\begin{lem}
\label{lem:Exchange order}The feasibility of \textbf{FP5 }is uninfluenced
by separating all $\theta_{m,k}^{j}$ from the procedure of calculating
a cumulative average in Eq. (\ref{eq:28-2}), i.e. the feasibility
remains unchanged by transforming (\ref{eq:28-2}) into the following
expression: 

\vspace{-15pt}

\end{lem}
\begin{equation}
\bar{P}_{m,k,j}^{{\rm {EH}}}=\frac{\sum_{i=1}^{j}\Psi_{k,i}}{j}{\tilde{\theta}_{m,k}^{j}},\label{eq:48}
\end{equation}
\emph{where }${\tilde{\theta}_{m,k}^{j}=\frac{\sum_{i=1}^{j}\Psi_{k,i}\theta_{m,k}^{i}}{\sum_{i=1}^{j}\Psi_{k,i}}}$.\emph{
Moreover, the new variable $\left[\tilde{\theta}_{m,k}^{j}\right]$
also satisfies the same constraints for $\left[\theta_{m,k}^{j}\right]$
in }(\ref{eq:7-1-1}) and (\ref{eq:35}).\emph{ }
\begin{IEEEproof}
The equivalence can be verified by utilizing the relationship between
$\left[\theta_{m,k}^{j}\right]$ and $\left[\tilde{\theta}_{m,k}^{j}\right]$
to evaluate (\ref{eq:7-1-1}), (\ref{eq:35}) and (\ref{eq:28-2}).
\end{IEEEproof}
Lemma \ref{lem:Exchange order} can decouple the unknown matrix $\left[\theta_{m,k}^{j}\right]$
with a large size. In this case, we only need to calculate the cumulative
average EH rate of the $k$-th parent relay once, instead of calculating
it for each child relay (i.e. the $\left(m,k\right)$-th child relay
for all $m$) in \textbf{FP5}. Particularly, the cumulative average
EH rate of the $k$-th relay is $\bar{P}_{k,j}^{{\rm {EH}}}=\frac{\sum_{i=1}^{j}\Psi_{k,i}}{j}.$ 

With the help of the two lemmas, we can obtain a new problem whose
feasibility is equivalent to \textbf{FP5}, while is much easier to
check. The results are in the following Corollary \ref{cor:Equivalent Problem},
which can be directly proved utilizing Theorem \ref{theo: Feasibility condition}
and Lemmas \ref{lem:Fix theta} and \ref{lem:Exchange order}. 
\begin{cor}
\label{cor:Equivalent Problem}The feasibility of \textbf{FP5} is
equivalent to the feasibility of the following problem for all \textup{$j\in\mathcal{J}$}: 

\vspace{-15pt}

\emph{
\begin{align}
\textrm{\textrm{\textbf{FP6:}}}\nonumber \\
\textrm{find} & \phantom{=}\phi_{m,k},\,\tilde{\theta}_{m,k}^{j},\,\forall k\in\mathcal{K},\: m\in\mathcal{M}\nonumber \\
\textrm{s.t.} & \phantom{=}{\sum_{k\in{{\cal S}_{\eta,m}}}}\frac{{2\tilde{\theta}_{m,k}^{j}\bar{P}_{k,j}^{{\rm {EH}}}jN^{\textrm{C}}+\frac{2{\phi_{m,k}}E_{k}^{\textrm{init}}}{T^{\textrm{C}}}-2{{\hat{P}}_{m,k}}\left(\eta\right)}}{{jN^{\textrm{C}}{{\hat{P}}_{m,k}}\left(\eta\right)}}\ge1,\nonumber \\
 & \phantom{=}\:\:\forall\: m\in\mathcal{M},\label{eq:37}\\
 & \phantom{=}0\leq\phi_{m,k}\leq1,\:0\leq\tilde{\theta}_{m,k}^{j}\leq1,\:\forall m\in\mathcal{M},\: k\in\mathcal{K},\label{eq:38}\\
 & \phantom{=}{\sum\limits _{m=1}^{M}{\phi_{m,k}}=1,}\:\sum\limits _{m=1}^{M}\tilde{\theta}_{m,k}^{j}=1,\:\forall k\in\mathcal{K}.\label{eq:40}
\end{align}
}
\end{cor}

\begin{rem}
We have obtained a highly simplified version of the feasibility problem
from the challenging original one while most steps are equivalent,
except for the approximation accompanied by employing Theorem \ref{theo: Feasibility condition}.
For \textbf{FP6}, it is a linear programming problem, with the complexity
linear with $\sqrt{N^{E}}MK$. Moreover, the variable size is greatly
reduced. We just need to solve \textbf{FP6} for all $j\in\mathcal{J}$
separately, and check if \textbf{FP6} is feasible for all $j\in\mathcal{J}$. 
\end{rem}
\vspace{-15pt}

\subsection{Proposed Power Assignment and Relay Selection Algorithm}

So far, the feasibility problem for the epigraph form of problem \textbf{OP1}
with a given $\eta$ is obtained. The following problem can provide
a sub-optimal solution for \textbf{OP1}

\vspace{-15pt}

\begin{align*}
\textrm{\textrm{\textbf{OP3:}}}\\
{\mathop{\min}\limits _{\left[\phi_{m,k}\right],\,\left[\tilde{\theta}_{m,k}^{j}\right],\,\eta}} & \phantom{=}\eta\\
\textrm{s.t.} & \phantom{=}{\textrm{Constraints }\eqref{eq:37}\sim\eqref{eq:40}.}
\end{align*}
Specifically, if \textbf{OP3} is feasible, then \textbf{OP1} is feasible,
and the optimal value of \textbf{OP3} provides a performance lower
bound to \textbf{OP1}. According to the two-phase pattern of the feasibility
for \textbf{OP3}, we can use a bisection method to find its optimal
solution, which provides a sub-optimal solution for \textbf{OP1}.
The proposed algorithm is shown as Algorithm \ref{alg: Bisec_Multiple},
where $\epsilon$ represents the required accuracy. 

\begin{algorithm}
\textbf{Initialization:}

\ \ \ \ Obtain $\eta_{m,k}^{\textrm{L}}$ by solving $U_{th}=U_{m,k}\left(\eta_{m,k}^{\textrm{L}},P_{k,\textrm{max}}^{{\rm {tr}}}\right)$.
If no feasible solution\FeasibleFootnote, set $\eta_{m,k}^{\textrm{L}}=0$.

\ \ \ \ Obtain $\eta_{m,k,j}^{\textrm{U}}$ by solving $U_{th}=U_{m,k}\left(1/\eta_{m,k,j}^{\textrm{U}},\Psi_{k,j}\right)$.
If no feasible solution\FeasibleFootnote, set $\eta_{m,k,j}^{\textrm{U}}=0$.

\ \ \ \ Set $\eta^{\textrm{L}}\leftarrow\underset{k\in\mathcal{K},\, m\in\mathcal{M}}{\textrm{max}}\:\eta_{m,k}^{\textrm{L}}$,
$\tilde{\eta}^{\textrm{U}}\leftarrow\underset{k\in\mathcal{K}}{\textrm{min}}\:\underset{m\in\mathcal{M},\, j\in\mathcal{J}}{\textrm{max}}\:\eta_{m,k,j}^{\textrm{U}}$,
$\eta^{\textrm{U}}=1/\tilde{\eta}^{\textrm{U}}$. 

\textbf{While} $\eta^{\textrm{U}}-\eta^{\textrm{L}}>\epsilon$

\ \ \ \ $\eta^{\textrm{M}}\leftarrow\frac{1}{2}\left(\eta^{\textrm{U}}+\eta^{\textrm{L}}\right)$. 

\ \ \ \ Obtain ${\cal S}_{\eta,m}$, ${{\hat{P}}_{m,k}}\left(\eta\right)$,
$\forall m\in\mathcal{M},\: k\in\mathcal{K}$ by (\ref{eq:29-2})
and (\ref{eq:16-1}), respectively.

\ \ \ \ \textbf{If} problem \textbf{FP6 }is feasible for every
$j\in\mathcal{J}$

\ \ \ \ \ \ \ \ $\eta^{\textrm{U}}\leftarrow\eta^{\textrm{M}}$.

\ \ \ \ \textbf{Else}

\ \ \ \ \ \ \ \ $\eta^{\textrm{L}}\leftarrow\eta^{\textrm{M}}$.

\ \ \ \ \textbf{End if}

\textbf{End while}

\textbf{Result:} 

\ \ \ \ The desired sub-optimal $\eta$ is $\eta^{\textrm{U}}$;

\ \ \ \ the transmit power matrix is $\mathbf{P}=\left[P_{m,k}^{{\rm {tr}}}\right]$
with $P_{m,k}^{{\rm {tr}}}={{\hat{P}}_{m,k}}\left(\eta\right)$;

\ \ \ \ the relay pre-selection is $\mathcal{S}_{m}={\cal S}_{\eta,m}$;

\ \ \ \ in each transmission block, the $m$-th source selects
one active relay (according to Eq. (\ref{eq:2})) randomly from $\mathcal{S}_{m}$.

\protect\caption{\label{alg: Bisec_Multiple}\hspace{-0.0405in}\textbf{:} \hspace{0.0405in}Suboptimal
power assignment and relay selection for \textbf{OP1}. }
\end{algorithm}

In Algorithm \ref{alg: Bisec_Multiple}, we first obtain the upper
bound and lower bound for bisection. The lower bound for the source
transmit power $\eta$ is obtained by having all the relays transmit
with their peak power, while the upper bound is obtained by having
an arbitrary relay transmit with its EH rate. Next, during each loop
of the bisection, with a given $\eta$, we determine a temporary relay
pre-selection result $\mathcal{\mathcal{S}_{\eta}}$, and a temporary
power assignment matrix $\hat{\mathbf{P}}_{\eta}={{\hat{P}}_{m,k}}\left(\eta\right)$.
Then, the feasibility of \textbf{FP6} is checked for all $j\in\mathcal{J}$.
If it is feasible, we reduce the value of $\eta$; otherwise, we increase
it. The smallest feasible $\eta$ is the sub-optimal value of \textbf{OP1}.
Note that in the result of Algorithm \ref{alg: Bisec_Multiple}, relay
pre-selection and power assignment are fixed within the whole transmission
duration, while relay selection needs to be performed at the beginning
of each transmission block. 

\footnotetext{If the equation has no feasible solution, it means that some relays have too poor relay-destination channels. Consequently, even if the source uses an infinite transmit power, the required utility still cannot be satisfied. }
\begin{rem}
The proposed cooperation strategy with joint power assignment and
relay selection has a low computational complexity, and also a low
requirement on the side information. Particularly, only the statistical
channel side information is needed. Furthermore, when the relay number
is very large, the offline EH information can also be exempted. In
contrast, the previously proposed transmission strategies for EH networks
are not quite practical, such as the offline approach that requires
all the offline side information \citep{EH_DWF_time}, \citep{EH_Broad_time,EH_DWF_Mac,EH_interference},
\citep{EH_two_hop6}, or the online algorithm with a high computational
complexity \citep{Energy_allo_EH_cons,Outage_min_EH_DP}. Moreover,
as will be shown in the simulation, the performance of Algorithm \ref{alg: Bisec_Multiple}
is close to optimal. On the other hand, this approach is very general
in the sense that the development is independent of the actual utility
function expression, as long as it is monotonic. 
\end{rem}

\section{Simulation Results}

In this section, we will provide simulation results to demonstrate
the performance of the proposed cooperation strategy. As shown in
Fig. \ref{fig:system_model}, in the simulation we consider a rectangular
area with length $L_{x}$ and width $L_{y}$, respectively. Without
loss of generality, the bottom-left of the rectangle is set as the
origin. The sources and destinations are located on two opposite sides,
with the coordinates given as $\left(0,\frac{m}{M+1}L_{y}\right)$
for the $m$-th source, while $\left(L_{x},\frac{m}{M+1}L_{y}\right)$
for the $m$-th destination. We assume that all the relays are uniformly
distributed inside the rectangle. 

In particular, we consider both the free-space path loss and small
scale fading. Denote the reference distance for the path loss as $d_{0}=10\textrm{ m}$,
and the free-space path loss at $d_{0}$ for a carrier frequency 2.4
GHz is calculated as 60 dB based on $\left[\lambda/\left(4\pi d\right)\right]^{2}$,
where $\lambda$ is the wavelength. We set $L_{x}=L_{y}=10d_{0}$.
The small scale fading is Rayleigh fading, with scale parameter 1.
The channel bandwidth is $B=1\textrm{ M}\textrm{Hz}$, and noise power
spectral density is $N_{0}=10^{\text{\textminus}16}\textrm{ W}/\textrm{Hz}$,
while the transmission block length $T^{\textrm{C}}$ is $10\textrm{ ms}$.
Unless otherwise mentioned, we set the EH interval as $T^{\textrm{E}}=5T^{\textrm{C}}$,
and the total transmission duration as $T=5T^{\textrm{E}}$. For each
realization, the EH rate of each relay is drawn randomly with uniform
distribution in $[\bar{P}^{\textrm{EH}}(1-\alpha),\:\bar{P}^{\textrm{EH}}(1+\alpha)]$,
where $\bar{P}^{\textrm{EH}}$ is the average EH rate, and $0<\alpha<1$.
In the simulation, the default value of $\alpha$ is 0.5, while the
maximum transmit power for each relay and the average source transmit
power are set as $P_{k,\textrm{max}}^{{\rm {tr}}}=2\textrm{ W}$ and
$\bar{P}^{\textrm{EH}}=20\textrm{ mW}$, respectively. 

\vspace{-15pt}

\subsection{The Single S-D Pair Case}

\begin{figure}
\begin{centering}
\includegraphics[scale=0.55]{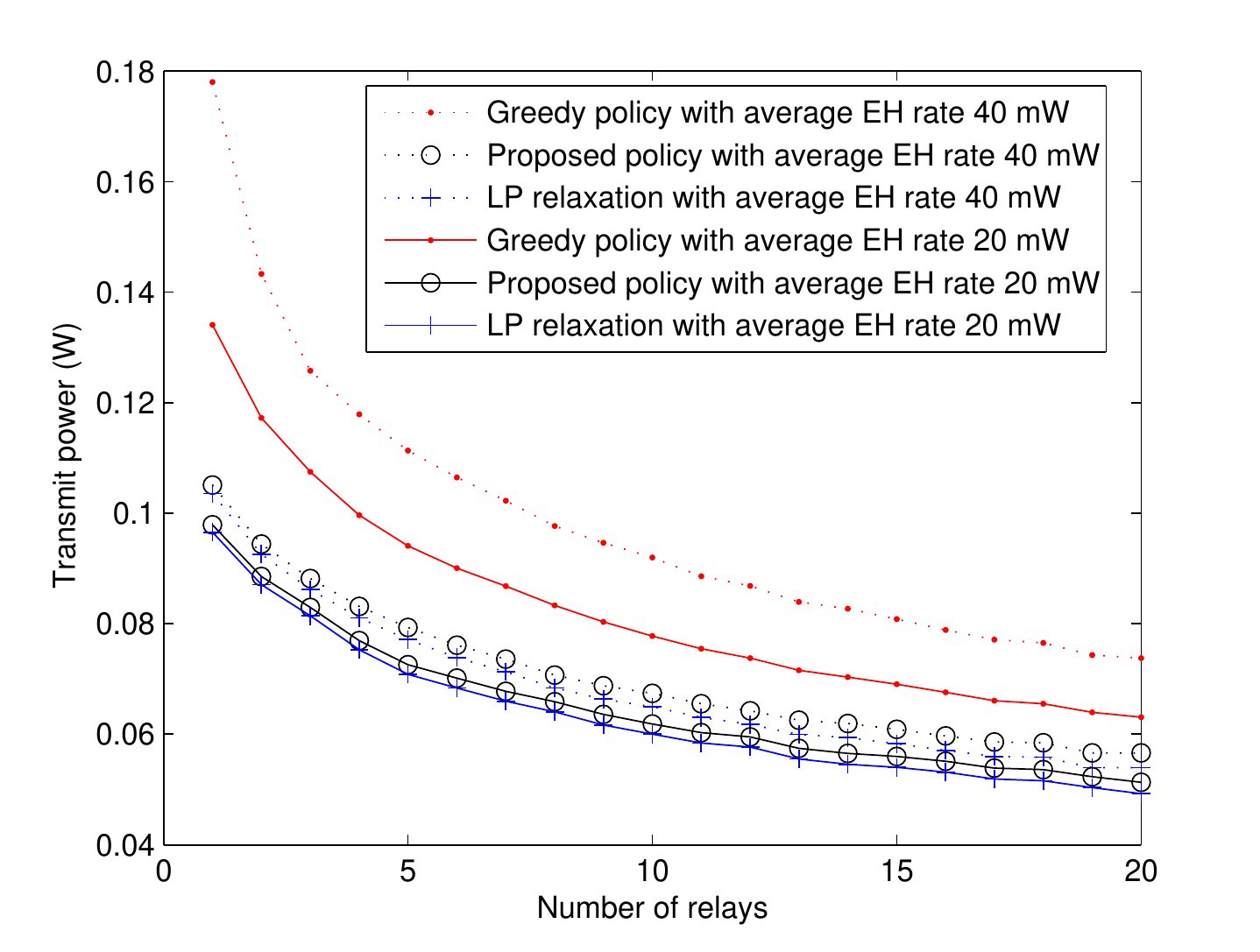}
\par\end{centering}

\protect\caption{\label{fig:Ptr_Relay_Single_1000p}Source transmit power versus the
relay number, with two average EH rates as 20 mW and 40 mW, respectively. }
\end{figure}

To demonstrate the advantage of the proposed  approach, we will compare
it with a \emph{greedy policy}, which will always select the relay
that can provide the best performance within each transmission block,
while the selected relay exhausts its energy. Such a policy does not
take the full EH constraints into consideration. We will also compare
with a performance upper bound, which, different from Algorithm \ref{alg: Bisec_Multiple},
solves the linear programming (LP) relaxation of \textbf{FP2 }instead
of using our proposed sufficient condition (\ref{eq:20}). In the
simulation, we adopt the successful transmission probability $\ensuremath{\mathbb{P}^{\textrm{s}}=1-\mathbb{P}^{\textrm{o}}}$
as the utility function, where $\mathbb{P}^{\textrm{o}}$ is the outage
probability, and the outage probability with a single AF relay can
be calculated based on Eq. (3.65) in \citep{Cooperative_com_book1}.
The SNR threshold is set as $\gamma=1$. The required successful transmission
probability is set as 0.99. The source transmit powers versus the
relay number for different policies are plotted in Fig. \ref{fig:Ptr_Relay_Single_1000p}.
As a reference, the transmit power of the direct link transmission
without the assistance of any relay is calculated as 0.995 W. We have
the following observations from the simulation results:
\begin{itemize}
\item As shown in Fig. \ref{fig:Ptr_Relay_Single_1000p}, the transmit power
of the proposed policy decreases with the relay number, and it is
much smaller than what is needed with the direct link transmission,
which is 0.995 W. The performance of the proposed algorithm is close
to that of the performance upper bound, which reveals the effectiveness
of the proposed methodology. 
\item The proposed policy outperforms the greedy policy which only considers
the current transmission block. For example, with average EH rate
20 mW, to have the source transmit power less than 0.08 W, the proposed
policy requires around 5 relays, while the best-effort policy needs
around 15 relays.
\item We see that increasing either the relay number or the average EH rate
can reduce the required source transmit power. Meanwhile, increasing
the relay number has a more obvious effect. The intuitive explanation
is that doubling the relay number will not only improve the total
available energy of all EH relays, but also reduce the (S-R or R-D)
transmission distances on average for each pair.
\end{itemize}

In the following, we will check the two asymptotic properties in Theorem
\ref{theo: Feasibility condition}, i.e. the necessity property and
the deterministic property. 

\begin{figure}
\begin{centering}
\includegraphics[scale=0.55]{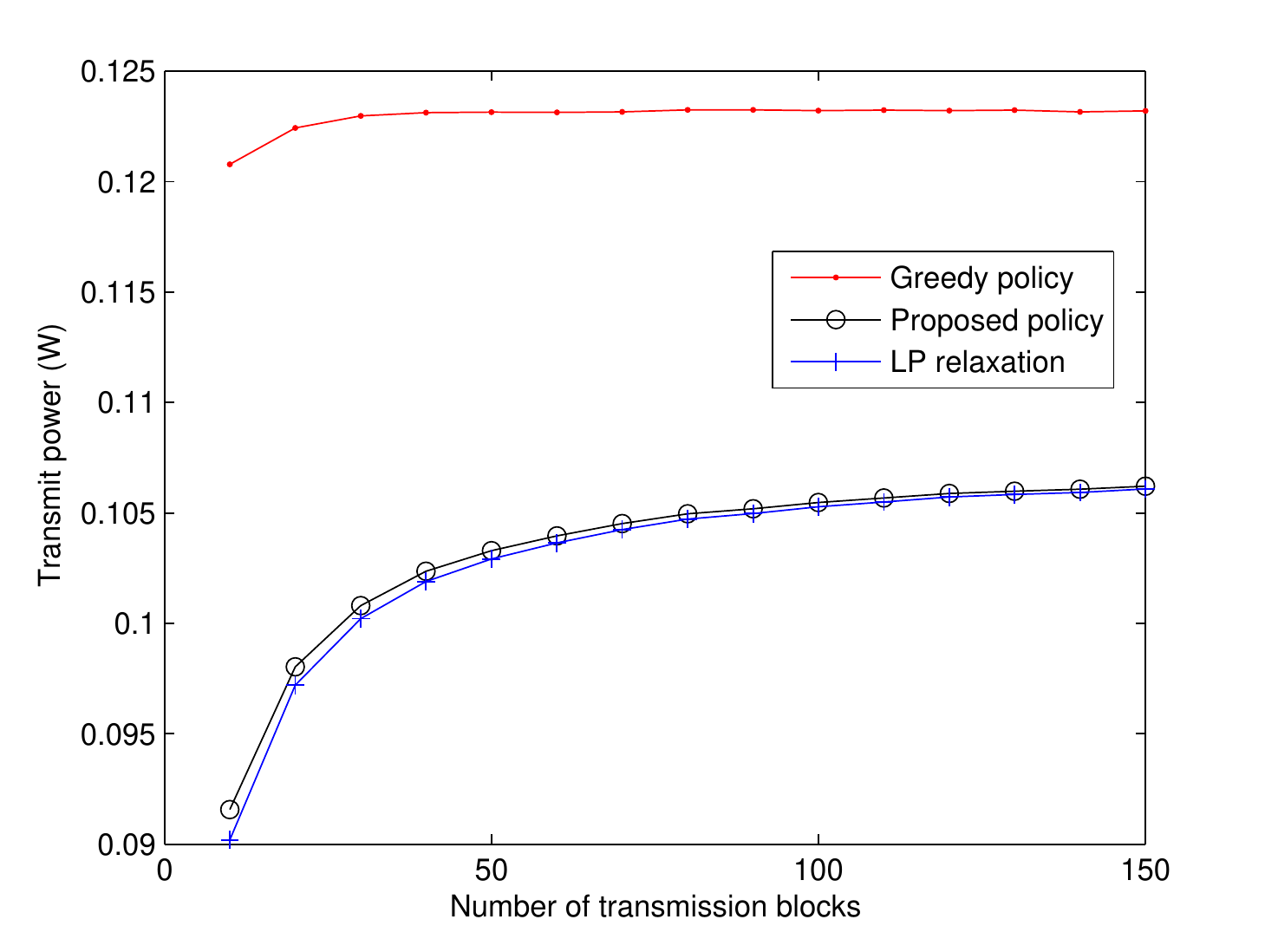}
\par\end{centering}

\protect\caption{\label{fig:Converge_AsymBlock}Source transmit power versus the number
of transmission blocks with 5 relays.}
\end{figure}

For the necessity property, the source transmit power versus transmission
block numbers with 5 relays are shown in Fig. \ref{fig:Converge_AsymBlock}.
We can see that even when there are only 10 transmission blocks in
each EH interval, the gap with the performance upper bound is already
small. When there are more than 100 transmission blocks, the difference
becomes indistinguishable from the figure. As in practical systems,
for the EH resources such as solar energy \citep{Outage_min_EH_DP,Predict_EH},
the EH interval (many seconds to even several minutes) are normally
much larger than the transmission block length (milliseconds to several
hundred milliseconds). In such cases, the performance of our policy
will be quite close to optimal. Note that, with the increase of transmission
blocks, the performances of all curves degrades gradually, which is
due to our fixed initial energy value setting, and a larger number
of transmission blocks means less available energy on average. 

\begin{figure}
\begin{centering}
\subfloat[The average transmit power for the under-estimation case versus the
number of relays.]{\begin{centering}
\includegraphics[scale=0.54]{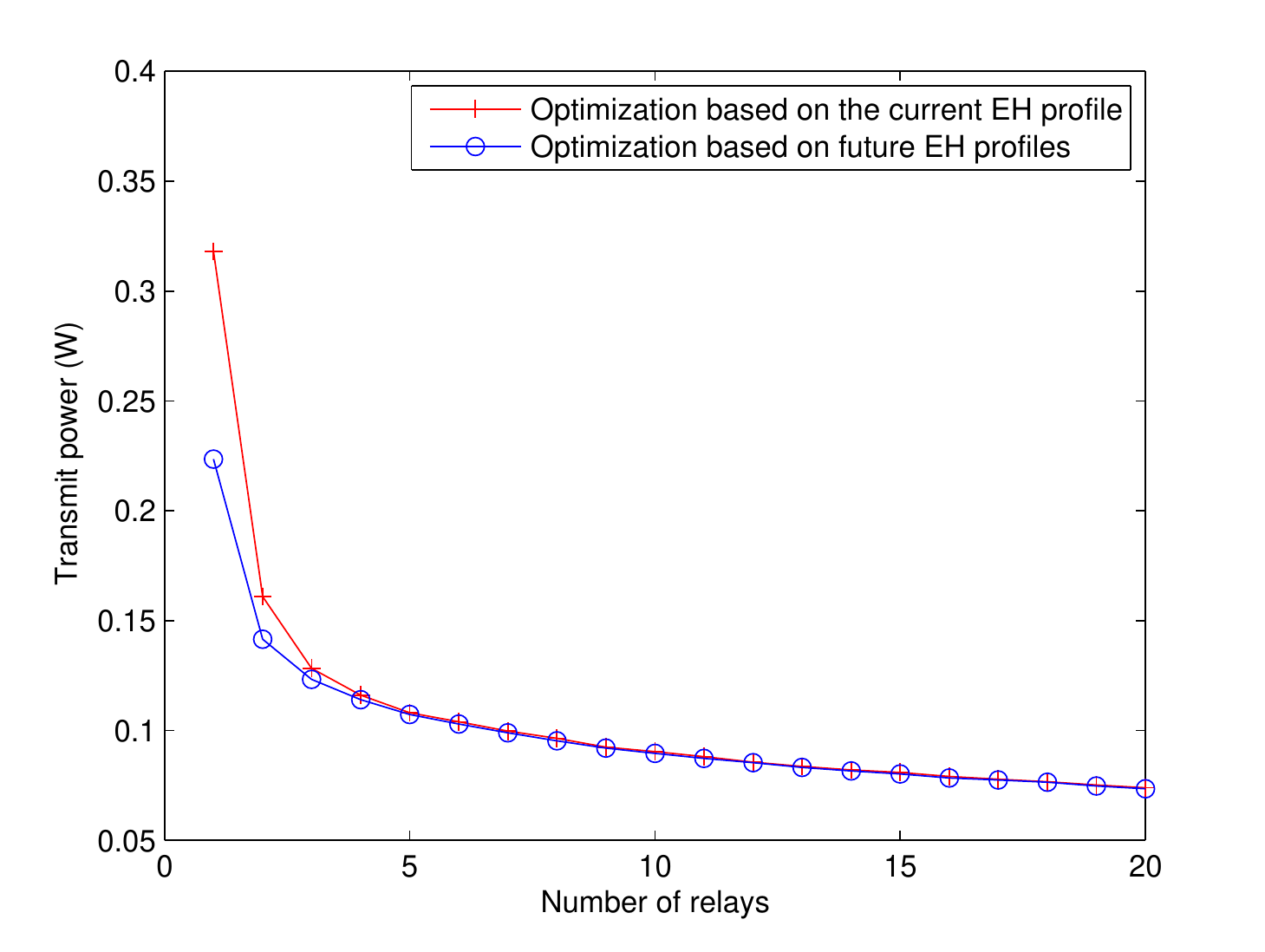}
\par\end{centering}

}
\par\end{centering}

\begin{centering}
\subfloat[The average ratio of no energy outage for the policy using only current
EH rates versus the number of relays.]{\begin{centering}
\includegraphics[scale=0.54]{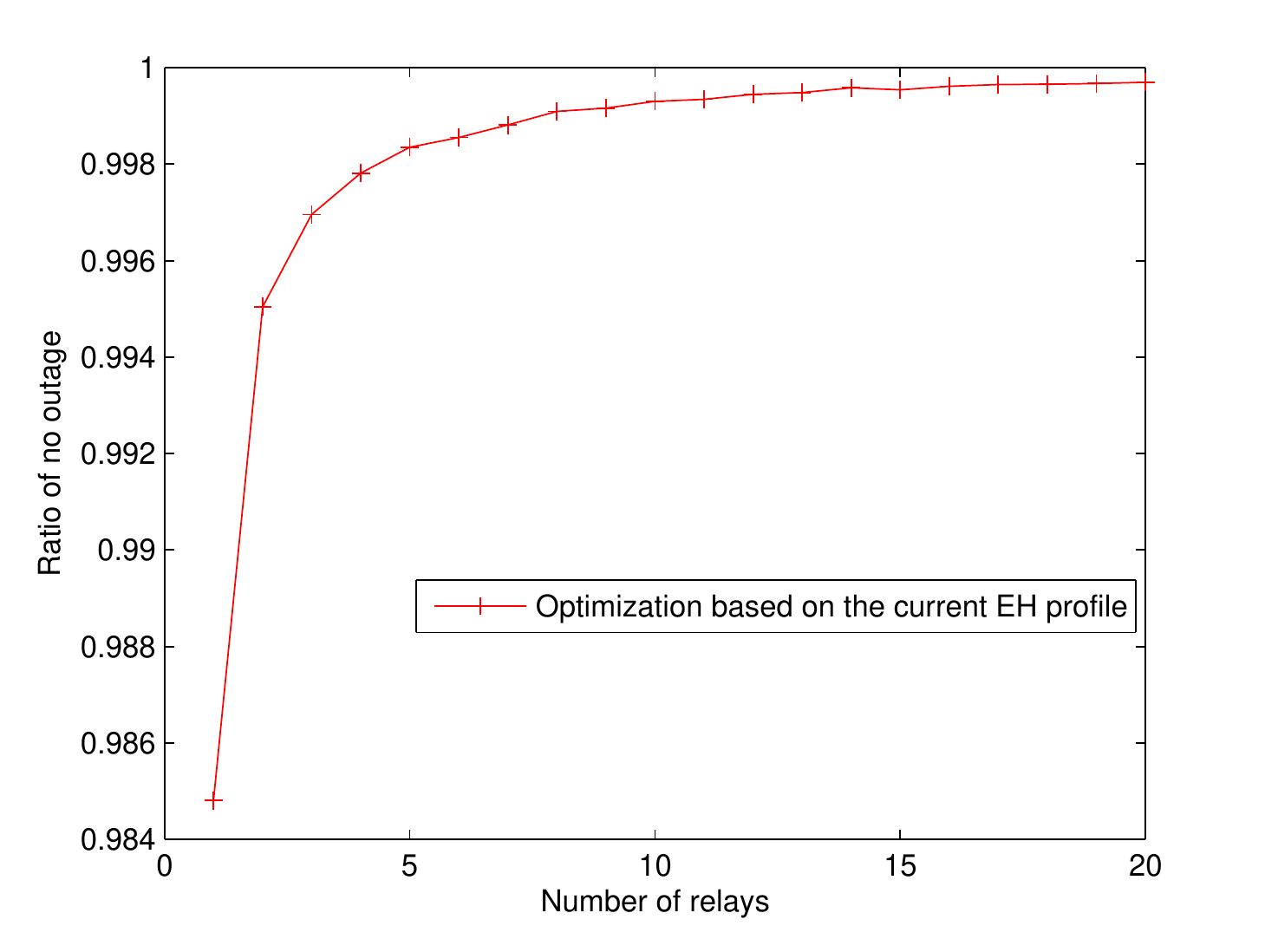}
\par\end{centering}

}
\par\end{centering}

\protect\caption{\label{fig:Converge_AsymRelay}Performance loss of using only the
current EH profiles. }
\end{figure}

Next, we check the deterministic property. Without the offline energy
side information, due to the randomness of future energy arrivals,
inevitably we can no longer deterministically guarantee all the constraints
are satisfied. In general, some performance loss will happen. In particular,
since we decide the transmit power/relay pre-selection based only
on the current side information, it is possible that we under-estimate
or over-estimate the future relay energy harvesting condition. Naturally,
we may set a too high or too low source transmit power, and this will
possibly then bring about higher source power consumption, or energy
outage, respectively. In the following, these two kinds of performance
loss will be discussed separately, and shown to become negligible
when increasing the relay number. \emph{In the case of under-estimation
of future relay EH situation}, from the second EH interval, there
will be some energy waste for the source. The average transmit power
for this under-estimation case%
\footnote{When calculating the average transmit power in the following simulation,
we only take into account the realizations where we have under-estimated
the future energy arrival and assign a too high power to the source.%
} with $N^{\textrm{C}}=1000$ is shown in Fig. \ref{fig:Converge_AsymRelay}
(a). We can see that when the relay number is larger than 6, the gap
is already very small. \emph{For the case of over-estimation of future
relay EH situation}, we may assign the source a too low transmit power,
and thus in some future EH intervals, the required QoS cannot be guaranteed
deterministically, which is called the \emph{energy outage}. In the
simulation, we count the respective ratio of transmission blocks not
suffering energy outage. The average ratio of no energy outage for
the policy using only current EH rates with $N^{\textrm{C}}=1000$
is shown in Fig. \ref{fig:Converge_AsymRelay} (b). We can see that
this ratio is no smaller than 0.985, and is approaching 1 with the
increase of the relay number.

\vspace{-15pt}

\subsection{The Multiple S-D Pair Case}

\begin{figure}
\begin{centering}
\subfloat[Source transmit power versus the number of relays. ]{\begin{centering}
\includegraphics[scale=0.54]{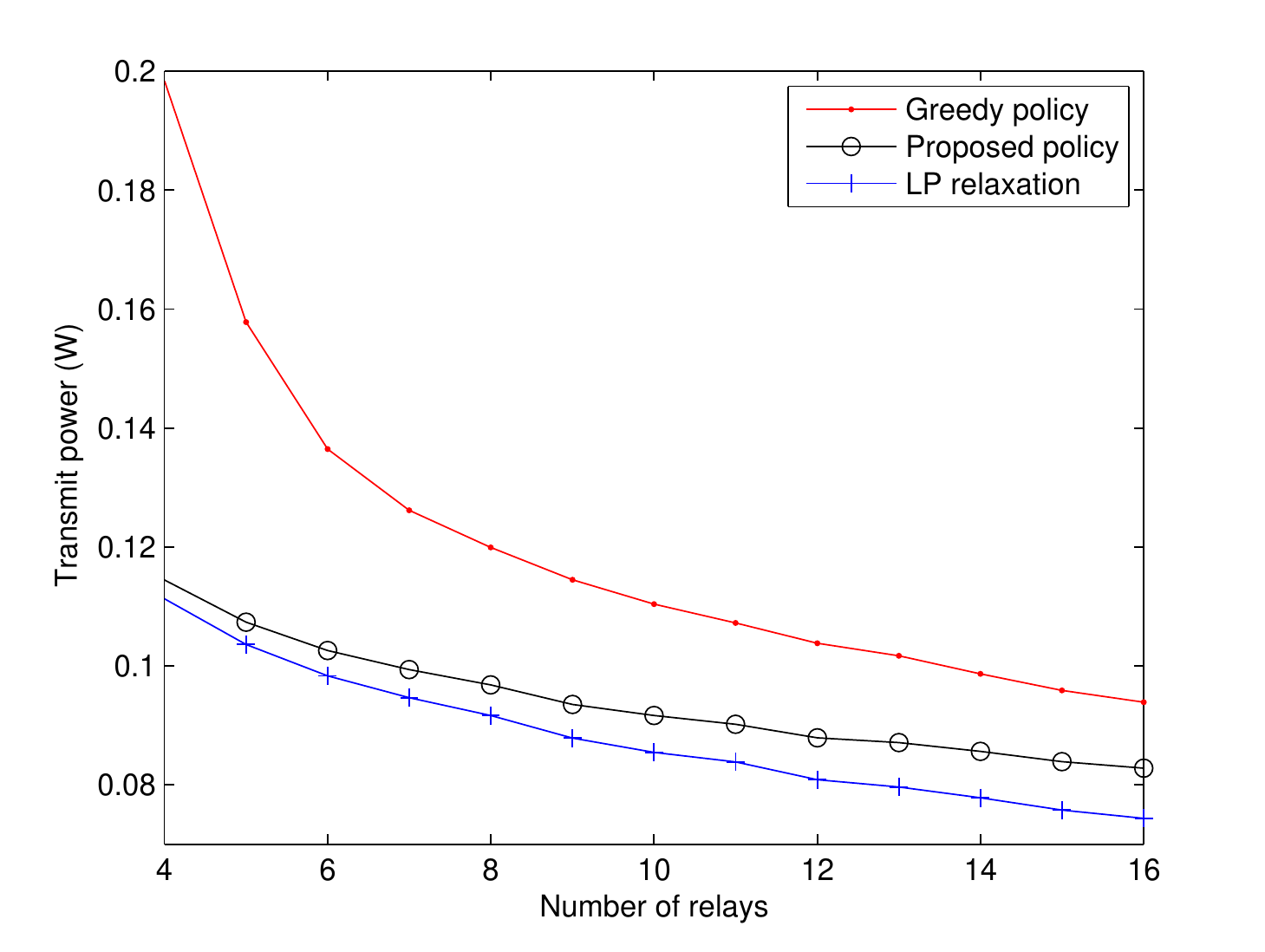}
\par\end{centering}

}
\par\end{centering}

\begin{centering}
\subfloat[Source transmit power versus the number of S-D pairs.]{\begin{centering}
\includegraphics[scale=0.54]{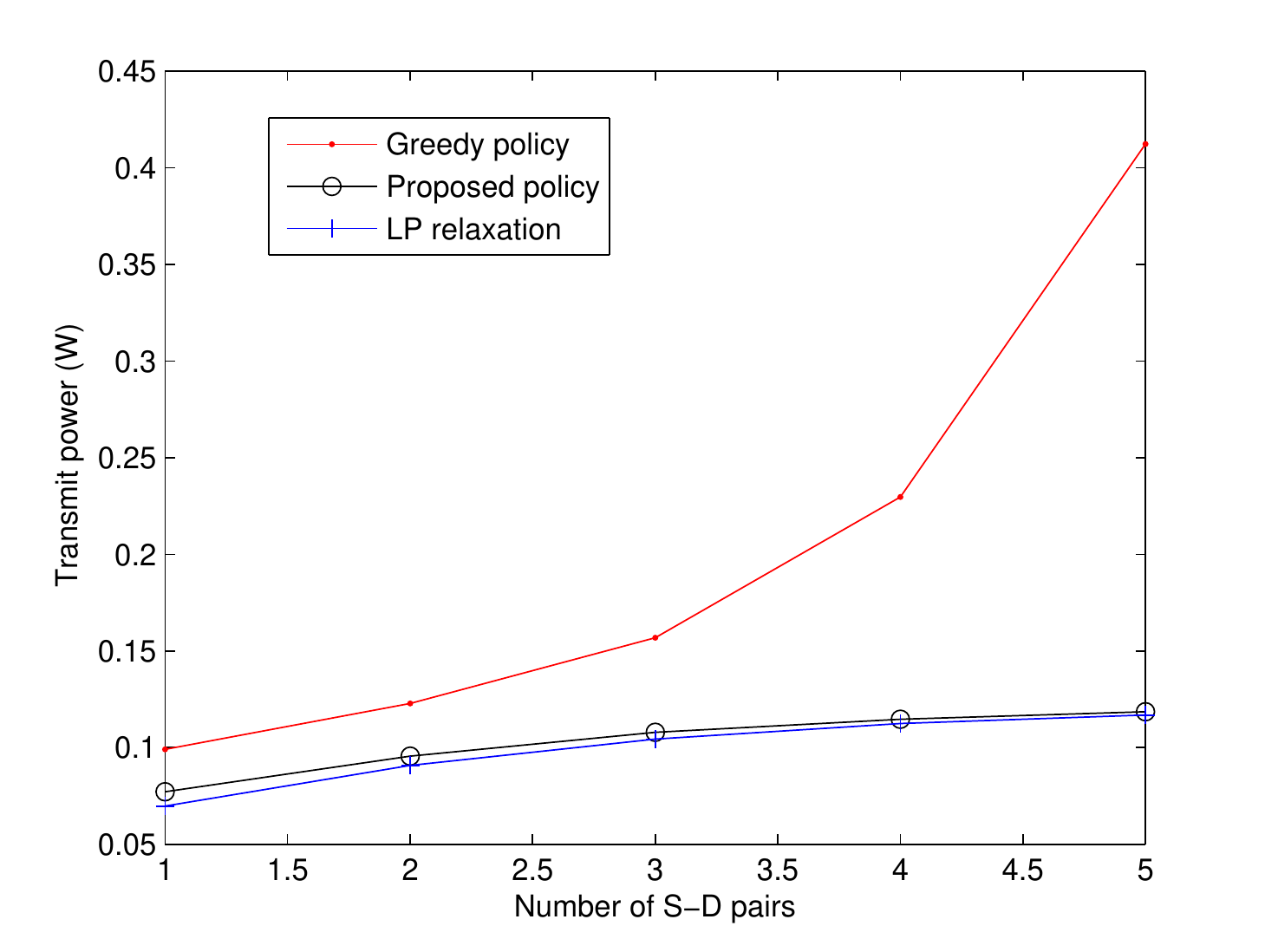}
\par\end{centering}

}
\par\end{centering}

\protect\caption{\label{fig:Ptr_Relay_Multi}Source transmit power for the multi-pair
case.}
\end{figure}

For the general case with multiple S-D pairs, we will also make a
comparison with a \emph{greedy policy}, for which, in each transmission
block, relay selection is executed to minimize the instantaneous source
transmit powers sequentially for all the S-D pairs in the current
transmission block. On the other hand, we compare both policies with
a performance upper bound, for which the feasibility is checked by
solving the LP relaxation of \textbf{OP1 }instead of solving \textbf{OP1}
with Algorithm \ref{alg: Bisec_Multiple}. The source transmit powers
for different policies versus the number of relays with 3 S-D pairs,
and versus the number of S-D pairs with 5 relays are plotted in the
Fig. \ref{fig:Ptr_Relay_Multi}. We have the following observations
from the simulation results:
\begin{itemize}
\item As shown in Fig. \ref{fig:Ptr_Relay_Multi} (a), for the multi-pair
system, the transmit powers also decrease with the number of relays,
and to have the same source transmit power more relays will be needed
than the single-pair case. The performance of the proposed policy
is quite close to that of the LP relaxation, which demonstrates the
effectiveness of Algorithm \ref{alg: Bisec_Multiple}. 
\item The greedy policy has a performance loss compared to the proposed
policy, as it only takes the current transmission block into consideration.
For example, to have the source transmit power less than 0.1 W, the
proposed policy requires more than 6 relays, while the greedy policy
requires more than 13 relays.
\item From Fig. \ref{fig:Ptr_Relay_Multi} (b), as the number of S-D pairs
increases, both the proposed policy and the greedy policy will experience
performance degradation, as on average the number of relays that can
assist each pair decreases. Meanwhile, the gap between the two policies
increases with the number of S-D pairs. Thus, the proposed policy
is more advantageous with a large number of S-D pairs. 
\end{itemize}

\section{Conclusions and Future works}

In this work, we proposed to achieve energy savings for a conventional
wireless network by providing energy diversity via multiple EH relays.
The proposed cooperation strategy is of low complexity and only depends
on the statistical channel side information. Particularly, an energy
hardening effect was revealed, which has demonstrated that it is possible
to relax the requirement of the offline EH information, and this effect
hence greatly simplifies the system design. Simulation results have
demonstrated that such a simple cooperation strategy can overcome
the low EH rate at each single EH relay, and provide significant power
gain to improve the source-destination communication. For the future
work, it would be interesting to apply the proposed algorithms to
other EH networks.

\section*{Appendix \ \ \ \ \ \ \ Proof for Theorem \ref{theo: Feasibility condition}}

The proof is divided into three parts: The sufficiency of Eq. (\ref{eq:20}),
the necessity property, and the deterministic property.

\textbf{1. Sufficiency of Eq. (\ref{eq:20}).}

To show the feasibility of \textbf{FP3}, we need to show that with
(\ref{eq:20}) satisfied, we can always find a feasible solution $\boldsymbol{\textrm{Z}}=\left[z_{k,n}\right]$
such that all the constraints of \textbf{FP3} are satisfied. All the
relays are active for the first transmission block due to the initial
energy setting. Thus, we only need to prove the existence of a feasible
relay selection result for the $\left(L+1\right)$-th transmission
block, with $1\leq L\leq N-1$. Moreover, the $L$-th transmission
block is in the $l$-th EH interval, with $l=\left\lceil L/N^{C}\right\rceil $.
For convenience, denote the number of transmission blocks when the
$k$-th relay is selected during the first $L$ transmission blocks
as $n_{k,L}$. For the first $L$ transmission blocks, we have $\sum\limits _{k\in\mathcal{S}_{\eta}}n_{k,L}\leq L$,
where the equality is achieved if there exist active relays (i.e.
satisfying Eq. (\ref{eq:2})) for all these $L$ transmission blocks.
Furthermore, based on (\ref{eq:20}), $\forall l$ such that $1\leq l\leq N^{\textrm{E}}$,
we can verify

\vspace{-15pt}
\begin{equation}
\sum_{k\in{{\cal S}_{\eta}}}\left(\frac{\sum_{i=1}^{l}\Psi_{k,i}}{{{{\hat{P}}_{k}}\left(\eta\right)}}+\frac{E_{k}^{{\rm {init}}}/T^{\textrm{C}}-{{\hat{P}}_{k}}\left(\eta\right)/2}{{{{\hat{P}}_{k}}\left(\eta\right)}N^{\textrm{C}}}\right)\geq\frac{l}{2},\label{eq:21}
\end{equation}
and

\vspace{-15pt}
\begin{equation}
\sum_{k\in{{\cal S}_{\eta}}}\left(\frac{\sum_{i=1}^{l+1}\Psi_{k,i}}{{{{\hat{P}}_{k}}\left(\eta\right)}}+\frac{E_{k}^{{\rm {init}}}/T^{\textrm{C}}-{{\hat{P}}_{k}}\left(\eta\right)/2}{{{{\hat{P}}_{k}}\left(\eta\right)}N^{\textrm{C}}}\right)\geq\frac{l+1}{2}.\label{eq:22}
\end{equation}
Now, denote $x=\frac{L-lN^{\textrm{C}}}{N^{\textrm{C}}},$ and $y=1-x,$
then by calculating $x\textrm{ Eq. }\eqref{eq:21}+y\textrm{ Eq. }\eqref{eq:22}$,
and adopting Eq. (\ref{eq:1-2}), we have $\sum_{k\in{{\cal S}_{\eta}}}\left(2\sum_{n=1}^{L}\frac{P_{k,n}^{\textrm{EH}}}{{{{\hat{P}}_{k}}\left(\eta\right)}}\right.$
$\left.+2\frac{E_{k}^{{\rm {init}}}/T^{\textrm{C}}-{{\hat{P}}_{k}}\left(\eta\right)/2}{{{{\hat{P}}_{k}}\left(\eta\right)}}\right)\geq L\geq\sum\limits _{k\in\mathcal{S}_{\eta}}n_{k,L}.$
That is, $\sum\limits _{k\in\mathcal{S}_{\eta}}\left(2\frac{\sum_{n=1}^{L}P_{k,n}^{\textrm{EH}}+E_{k}^{{\rm {init}}}/T^{\textrm{C}}-{{\hat{P}}_{k}}\left(\eta\right)/2}{{{{\hat{P}}_{k}}\left(\eta\right)}}-n_{k,L}\right)\geq0.$
Following the pigeonhole principle, we can verify that there exists
at least one $\hat{k}\in\mathcal{S}_{\eta}$, such that $2\frac{\sum_{n=1}^{L}P_{\hat{k},n}^{\textrm{EH}}+E_{\hat{k}}^{{\rm {init}}}/T^{\textrm{C}}-{{\hat{P}}_{\hat{k}}}\left(\eta\right)/2}{{{{\hat{P}}_{\hat{k}}}\left(\eta\right)}}-n_{\hat{k},L}\geq0,$
i.e. $E_{\hat{k}}^{{\rm {init}}}+\sum_{n=1}^{L}P_{\hat{k},n}^{\textrm{EH}}T^{\textrm{C}}-{{{\hat{P}}_{\hat{k}}}\left(\eta\right)}n_{\hat{k},L}\frac{T^{\textrm{C}}}{2}\geq{{\hat{P}}_{\hat{k}}}\left(\eta\right)\frac{T^{\textrm{C}}}{2}.$
This means that the remaining energy of the $\hat{k}$-th relay after
$L$ transmission blocks is enough to support its assigned transmit
power. That is, it is active based on Eq. (\ref{eq:2}) and ready
for the relay selection of the $\left(L+1\right)$-th transmission
block. We can then set $z_{\hat{k},L+1}=1$ while $z_{k,L+1}=0$ for
$k\neq\hat{k},\: k\in\mathcal{K}$, for the $\left(L+1\right)$-th
transmission block. It can be verified that the obtained relay selection
result for $L\in\mathcal{N}$ satisfies all the constraints. Hence,
this is a feasible solution for \textbf{FP3}.

\textbf{2. Necessity property. }

It can be verified that the following inequality is a sufficient condition
for Eq. (\ref{eq:20}): $\zeta_{1}\left(\eta,\: l\right)={\sum_{k\in{{\cal S}_{\eta}}}}\frac{{2\bar{\Psi}_{k,l}}}{{{{\hat{P}}_{k}}\left(\eta\right)}}\geq1,\:\forall l.$
Therefore, we only provide the necessity of this inequality, instead
of Eq. (\ref{eq:20}), when $N^{\textrm{C}}\rightarrow\infty$. We
will do so by contradiction. Specifically, assume that there exists
one EH interval, e.g., the $l^{*}$-th, say, 
\begin{equation}
{\sum_{k\in{{\cal S}_{\eta}}}}\frac{{2\bar{\Psi}_{k,l^{*}}}}{{{{\hat{P}}_{k}}\left(\eta\right)}}<1,\label{eq:23}
\end{equation}
but the feasibility of \textbf{FP3} still holds. Similar to the proof
for the sufficiency property, we denote the number of blocks that
the $k$-th relay is selected during the first $L=l^{*}T^{{\rm {c}}}$
transmission blocks as $n_{k,L}$. Due to Eq. (\ref{eq:17})\textbf{
}of\textbf{ FP3}, there exist active relays for all the $L$ transmission
blocks. Thus, 
\begin{equation}
\sum\limits _{k\in\mathcal{S}_{\eta}}n_{k,L}=L.\label{eq:24}
\end{equation}
By jointly considering (\ref{eq:23}) and (\ref{eq:24}), we have
${\sum_{k\in{{\cal S}_{\eta}}}}\frac{{2\bar{\Psi}_{k,l^{*}}}}{{{{\hat{P}}_{k}}\left(\eta\right)}}L<L=\sum\limits _{k\in\mathcal{S}_{\eta}}n_{k,L},$
i.e. ${\sum_{k\in{{\cal S}_{\eta}}}}\left(\frac{{2\bar{\Psi}_{k,l^{*}}}L}{{{{\hat{P}}_{k}}\left(\eta\right)}}\right.$
$\left.-\hspace{-5pt}\phantom{\frac{{\bar{I}}}{{\hat{I}}}}n_{k,L}\right)<0.$
Following the pigeonhole principle, we can verify that there exists
at least one $\hat{k}\in\mathcal{S}_{\eta}$, such that $\frac{{2\bar{\Psi}_{\hat{k},l*}}L}{{{{\hat{P}}_{\hat{k}}}\left(\eta\right)}}-n_{\hat{k},L}<0,$
with $L=l^{*}N^{\textrm{C}}\rightarrow\infty$, $\frac{E_{\hat{k}}^{{\rm {init}}}+{2\bar{\Psi}_{\hat{k},l^{*}}}L}{{{{\hat{P}}_{\hat{k}}}\left(\eta\right)}}-n_{\hat{k},L}<0,$
i.e. $E_{\hat{k}}^{{\rm {init}}}+{2\bar{\Psi}_{\hat{k},l^{*}}}L-n_{\hat{k},L}{{{\hat{P}}_{\hat{k}}}\left(\eta\right)}<0.$
This violates the energy causality constraint (\ref{eq:6}). Therefore,
the feasibility of \textbf{FP3} is violated.

\textbf{3. Deterministic property.}

Denote $X_{k,l}=\boldsymbol{1}_{U_{k}\left(\eta,P_{k,\textrm{max}}^{\textrm{tr}}\right)\geq U_{\textrm{th}}}\frac{{2\bar{\Psi}_{k,l}lN^{\textrm{C}}+2E_{k}^{{\rm {init}}}/T^{\textrm{C}}-2{{\hat{P}}_{k}}\left(\eta\right)}}{lN^{\textrm{C}}}$
$\frac{K}{{{{\hat{P}}_{k}}\left(\eta\right)}},$ and $Y_{l}=\frac{\sum_{k=1}^{K}X_{k,l}}{K}$,
then we have $\zeta\left(\eta,\: l\right)=\frac{\sum_{j=1}^{l}Y_{j}}{l}.$
With a given $l$, and given locations, (i.e. given ${{{\hat{P}}_{k}}\left(\eta\right)}$),
$X_{k,l}$ with different $k$ are independent of each other, as all
the EH rates for different relays are independent. We can further
verify that all $X_{k,l}$ are with finite variances, as $\sigma_{X_{k,l}}^{2}\leq\left(E\left[X_{k,l}\right]\right)^{2}\leq\left(\frac{{2E\left[\bar{\Psi}_{k,l}\right]lN^{\textrm{C}}+2E_{k}^{{\rm {init}}}/T^{\textrm{C}}-2{{\hat{P}}_{k}}\left(\eta\right)}}{lN^{\textrm{C}}}\frac{K}{{{{\hat{P}}_{k}}\left(\eta\right)}}\right)^{2}\leq\left(\textrm{constant}\frac{K}{{{{\hat{P}}_{k}}\left(\eta\right)}}\right)^{2},$
while the finite value of $\frac{K}{{{{\hat{P}}_{k}}\left(\eta\right)}}$
can be verified non-trivially by the two-phase pattern for the feasibility
of \textbf{FP1}, as mentioned in the end of the last section. Then,
$\forall l$, $\sum_{k}\sigma_{X_{k,l}}^{2}/k^{2}<\infty$. Based
on the strong law of large numbers, cf. Theorem 3 in Chapter 2.9 in
\citep{Prob_survey}, $Y_{j}$ converges to be a deterministic value.
It can be further verified that $\zeta\left(\eta,l\right)=\frac{\sum_{j=1}^{l}Y_{j}}{l}$
will become deterministic.

{\footnotesize{}\bibliographystyle{IEEEtran}
\bibliography{example}
}{\footnotesize \par}

\begin{IEEEbiography}
[{\includegraphics[width=1in,height=1.25in,clip,keepaspectratio] {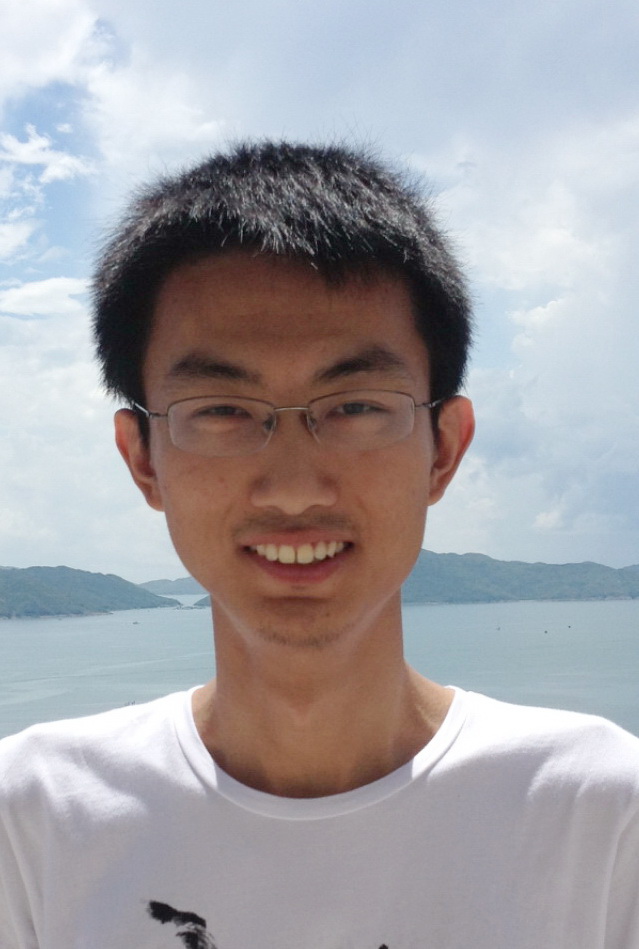}}]{Yaming Luo}
(S'11) received his B.Eng. degree in the Department of Communication Engineering from the Harbin Institute of Technology, Harbin, China, in 2010. He also obtained the honor of Outstanding Graduates of  Heilongjiang Province. He is currently working towards the Ph.D. degree in the  Department of Electronic and Computer Engineering at the Hong Kong University of  Science and Technology, under the supervision of Prof. Khaled B. Letaief. His current research interests include energy  harvesting networks, relay systems, and green communications.
\end{IEEEbiography}
\vspace*{0\baselineskip}
\begin{IEEEbiography}
[{\includegraphics[width=1in,height=1.25in,clip,keepaspectratio] {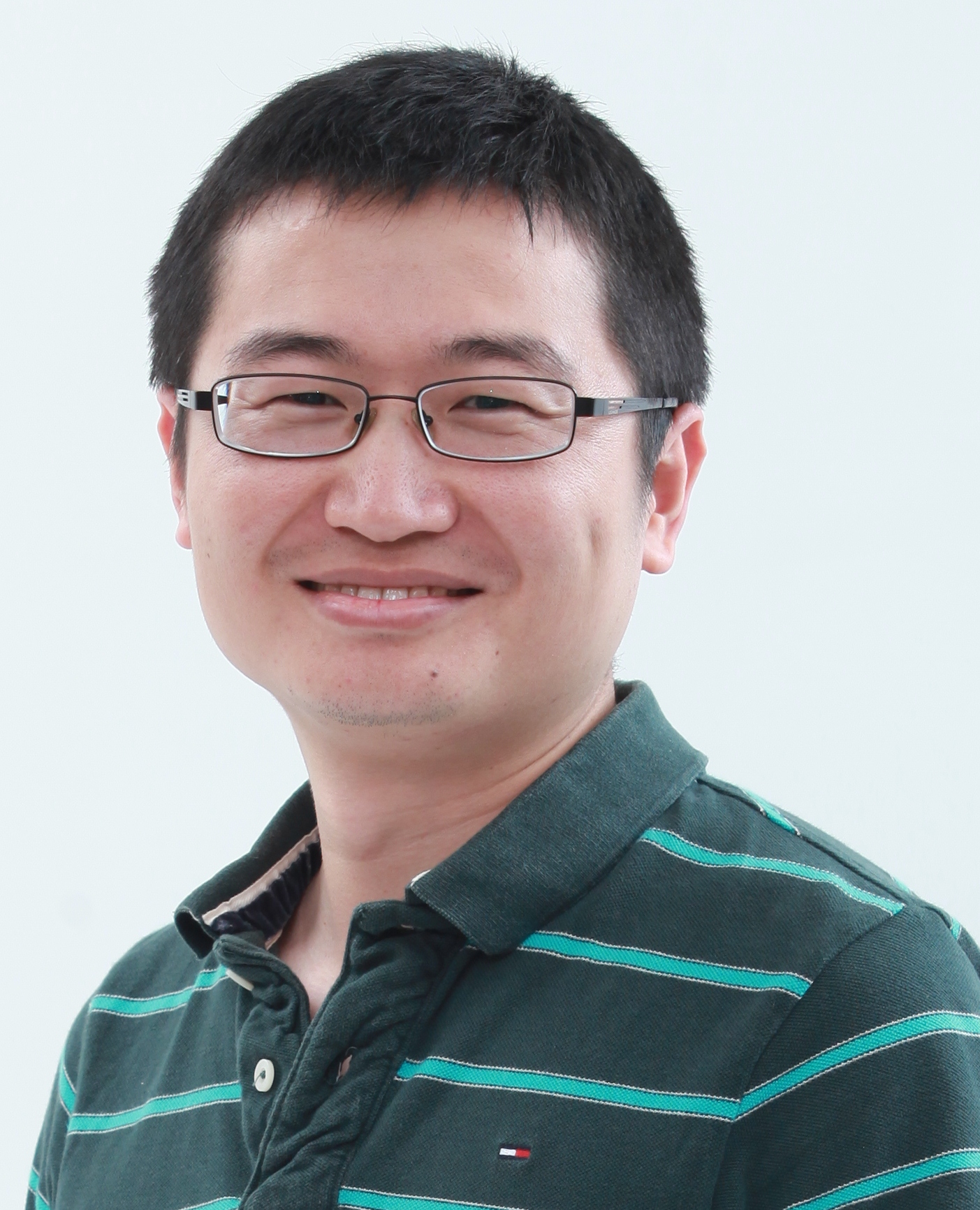}}]{Jun Zhang}
(S'06-M'10-SM'15) received the B.Eng. degree in Electronic Engineering from the University of Science and Technology of China in 2004, the M.Phil. degree in Information Engineering from the Chinese University of Hong Kong in 2006, and the Ph.D. degree in Electrical and Computer Engineering from the University of Texas at Austin in 2009. He is currently a Research Assistant Professor in the Department of Electronic and Computer Engineering at the Hong Kong University of Science and Technology (HKUST). Dr. Zhang co-authored the book Fundamentals of LTE (Prentice-Hall, 2010). He received the 2014 Best Paper Award for the EURASIP Journal on Advances in Signal Processing, and the PIMRC 2014 Best Paper Award. He is an Editor of IEEE Transactions on Wireless Communications, and served as a MAC track co-chair for IEEE WCNC 2011. His research interests include wireless communications and networking, green communications, and signal processing. 
\end{IEEEbiography}
\begin{IEEEbiography}
[{\includegraphics[width=1in,height=1.25in,clip,keepaspectratio] {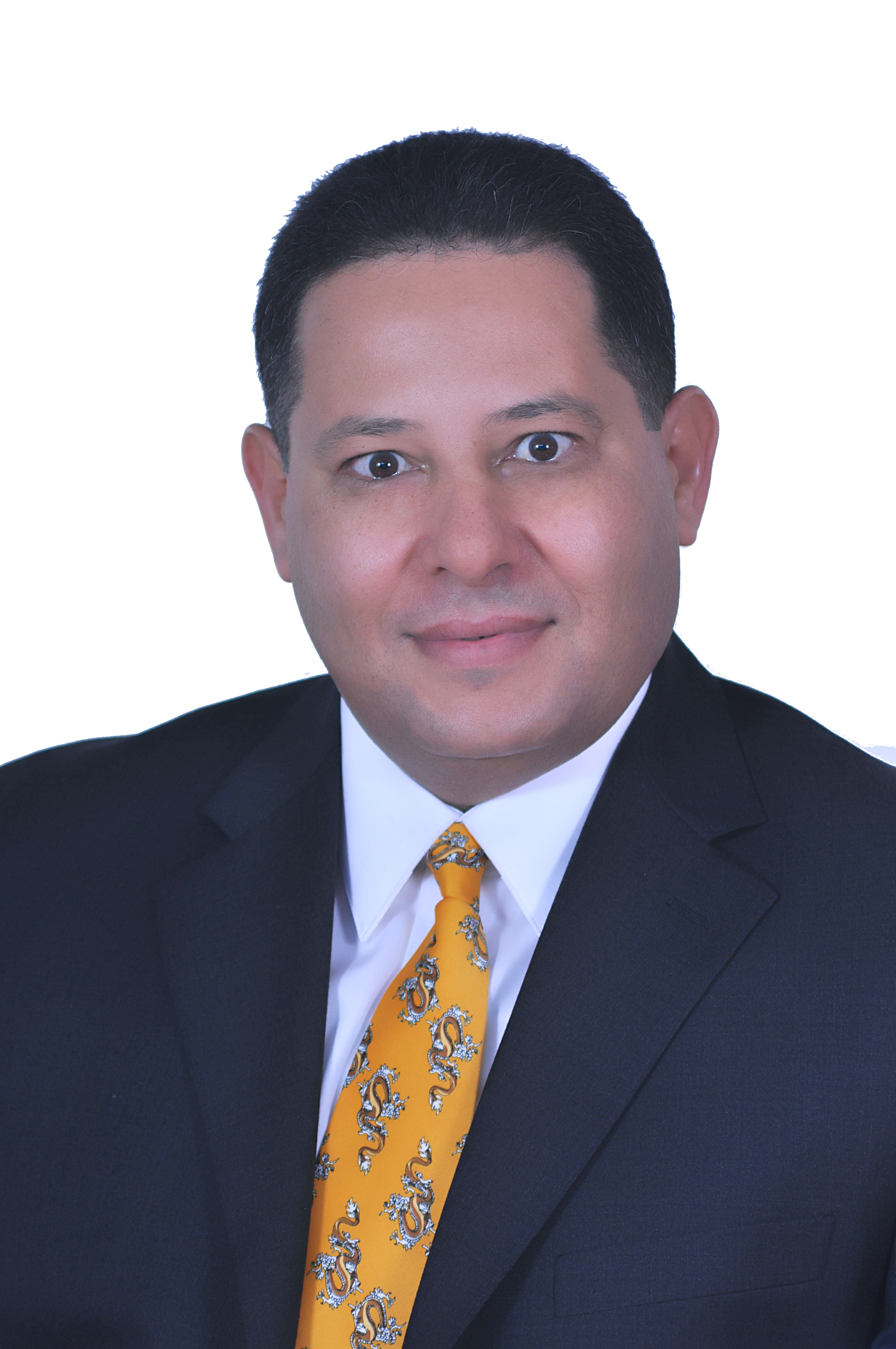}}]{Khaled B. Letaief}
(S'85-M'86-SM'97-F'03) received the BS degree \emph{with distinction} in Electrical Engineering (1984) from Purdue University, USA. He has also received the MS and Ph.D. Degrees in Electrical Engineering from Purdue University in 1986 and 1990, respectively.   
\par 
From 1990 to 1993, he was a faculty member at the University of Melbourne, Australia.  He has been with the Hong Kong University of Science and Technology.  While at HKUST, he has held numerous administrative positions, including the Head of the Electronic and Computer Engineering department, Director of the Center for Wireless IC Design, Director of Huawei Innovation Laboratory, and Director of the Hong Kong Telecom Institute of Information Technology.  He has also served as Chair Professor and Dean of HKUST School of Engineering.  Under his leadership, the School of Engineering has dazzled in international rankings (\emph{ranked \#14 in the world in 2015} according to QS World University Rankings).   
\par 
From September 2015, he joined HBKU as Provost to help establish a research-intensive university in Qatar in partnership with strategic partners that include Northwestern University, Carnegie Mellon University, Cornell, and Texas A\&M.
\par 
Dr. Letaief is a world-renowned leader in wireless communications and networks.  In these areas, he has over 500 journal and conference papers and given invited keynote talks as well as courses all over the world.  He has made 6 major contributions to IEEE Standards along with 13 patents including 11 US patents. 
\par 
He served as consultants for different organizations and is the founding Editor-in-Chief of the prestigious \emph{IEEE Transactions on Wireless Communications}.  He has served on the editorial board of other prestigious journals including the \emph{IEEE Journal on Selected Areas in Communications - Wireless Series} (as Editor-in-Chief).  He has been involved in organizing a number of major international conferences. 
\par 
Professor Letaief has been a dedicated educator committed to excellence in teaching and scholarship.  He received the \emph{Mangoon Teaching Award} from Purdue University in 1990; HKUST Engineering Teaching Excellence Award (4 times); and the Michael Gale Medal for Distinguished Teaching (\emph{Highest university-wide teaching award} at HKUST).  
\par 
He is also the recipient of many other distinguished awards including 2007 \emph{IEEE Joseph LoCicero Publications Exemplary Award}; 2009 IEEE Marconi Prize Award in Wireless Communications; 2010 Purdue University Outstanding Electrical and Computer Engineer Award; 2011 IEEE Harold Sobol Award; 2011 IEEE Wireless Communications Technical Committee Recognition Award; and 11 IEEE Best Paper Awards.   
\par 
He had the privilege to serve IEEE in many leadership positions including IEEE ComSoc Vice-President, IEEE ComSoc Director of Journals, and member of IEEE Publications Services and Products Board, IEEE ComSoc Board of Governors, IEEE TAB Periodicals Committee, and IEEE Fellow Committee.   
\par 
Dr. Letaief is a Fellow of IEEE and a Fellow of HKIE.  He is also recognized by Thomson Reuters as an \emph{ISI Highly Cited Researcher}. 
 
\end{IEEEbiography}

\end{document}